\documentclass[acmsmall,screen]{acmart}

\AtBeginDocument{%
  }

\setcopyright{rightsretained}
\acmDOI{10.1145/3656431}
\acmYear{2024}
\copyrightyear{2024}
\acmSubmissionID{pldi24main-p400-p}
\acmJournal{PACMPL}
\acmVolume{8}
\acmNumber{PLDI}
\acmArticle{201}
\acmMonth{6}
\received{2023-11-16}
\received[accepted]{2024-03-31}

\usepackage{listings}
\usepackage{syntaxdefs}
\usepackage{tikz}
\usepackage{tikz-cd}
\usepackage{stmaryrd}
\usetikzlibrary{positioning}
\usetikzlibrary{arrows}
\usetikzlibrary{calc}
\usetikzlibrary{automata}
\usepackage[ruled,vlined]{algorithm2e} 
\usepackage{wrapfig}
\usepackage{etoolbox}

\makeatletter
\patchcmd{\@algocf@start}
  {-1.5em}
  {0pt}
  {}{}
\makeatother

\begin{document}

\title{Linear Matching of JavaScript Regular Expressions}

\author{Aurèle Barrière}
\email{aurele.barriere@epfl.ch}
\orcid{0000-0002-7297-2170}
\author{Clément Pit-Claudel}
\email{clement.pit-claudel@epfl.ch}
\orcid{0000-0002-1900-3901}
\affiliation{%
  \institution{EPFL}
  \city{Lausanne}
  \country{Switzerland}}


\begin{abstract}

  Modern regex languages have strayed far from well-understood traditional regular expressions:
  they include features that fundamentally transform the matching problem.
  In exchange for these features, modern regex engines at times suffer from exponential complexity blowups, a frequent source of denial-of-service vulnerabilities in JavaScript applications.
  Worse, regex semantics differ across languages, and the impact of these divergences on algorithmic design and worst-case matching complexity has seldom been investigated.
  
  This paper provides a novel perspective on JavaScript's regex semantics by identifying a larger-than-previously-understood subset of the language that can be matched with linear time guarantees.
  In the process, we discover several cases where state-of-the-art algorithms were either wrong (semantically incorrect), inefficient (suffering from superlinear complexity) or excessively restrictive (assuming certain features could not be matched linearly).
  We introduce novel algorithms to restore correctness and linear complexity.
  We further advance the state-of-the-art in linear regex matching by presenting the first nonbacktracking algorithms for matching lookarounds in linear time:
  one supporting captureless lookbehinds in any regex language,
  and another leveraging a JavaScript property to support unrestricted lookaheads and lookbehinds.
  Finally, we describe new time and space complexity tradeoffs for regex engines.
  All of our algorithms are practical: we validated them in a prototype implementation, and some have also been merged in the V8 JavaScript implementation used in Chrome and Node.js.
  
\end{abstract}






\lstdefinestyle{rgx}{ %
  xleftmargin=2\parindent,
  captionpos=b,
  mathescape,
  basicstyle=\ttfamily,
  commentstyle=\itshape\color{MidnightBlue},
  identifierstyle=\color{black},
  keywordstyle=\bfseries,
  escapeinside={@}{@}
}

\lstdefinestyle{byt}{ %
  xleftmargin=2\parindent,
  captionpos=b,
  mathescape,
  basicstyle=\ttfamily,
  commentstyle=\itshape\color{MidnightBlue},
  identifierstyle=\color{black},
  keywordstyle=\bfseries,
}

\def\bytecode#1{\lstinline[style=byt]{#1}}
\def\tbytecode#1{\text{\bytecode{#1}}}

\def\rgx#1{\texttt{#1}}         
\def\regex#1{\texttt{/#1/}}     
\def\trgx#1{\text{\rgx{#1}}}
\def\tregex#1{\text{\regex{#1}}}
\def\noncap#1{\ensuremath{\llparenthesis}\,#1\,\ensuremath{\rrparenthesis}}
\def\eps{\ensuremath{\epsilon}}
\def\grp#1{\ensuremath{_{\##1}}}
\def\lkr#1{\ensuremath{_{\lessgtr#1}}}

\def\looknb#1{\ensuremath{\ell(#1)}}
\def\str#1{\texttt{"#1"}}
\def\group#1{\texttt{\##1}}
\def\looka#1{\ensuremath{\lessgtr\hspace{-0.1cm}#1}}
\def\undef{\texttt{undefined}}

\def\size#1{\ensuremath{|#1|}}
\def\psize#1{\ensuremath{||#1||}}
\def\bigo#1{\ensuremath{\mathcal{O}}\left(#1\right)}

\def\re{\ensuremath{\mathit{e}}}
\def\chara{\ensuremath{\mathit{c}}}
\def\quant{\ensuremath{\mathit{q}}}
\def\look{\ensuremath{\mathit{lk}}}
\def\instr{\ensuremath{\mathit{i}}}
\def\lbl{\ensuremath{\mathit{l}}}
\def\reg{\ensuremath{\mathit{reg}}}
\def\lid{\ensuremath{\mathit{lid}}}

\def\aurele#1{{\footnotesize\color{ACMBlue}{[A]}#1}}
\def\cpc#1{{\footnotesize\color{ACMRed}{[C]}#1}}

\newcommand{\parbf}[1]{\textbf{#1}}

\maketitle

\section{Introduction}

The expressive power and computational complexity of traditional regular expressions (composed only of characters, concatenations, alternations, and Kleene stars) are well understood.
From an expressive-power standpoint, they are known to be exactly equivalent to finite automata.  From a computational-complexity standpoint (deciding whether a regular expression $r$ matches a string $s$), well-known trade-offs exist between pre-processing and matching time.  For instance, one can create a deterministic finite automaton (DFA) for $r$ in worst-case time complexity $\bigo{2^{\size{r}}}$ and then perform matching in $\bigo{\size{s}}$, where $\size{r}$ and $\size{s}$ represent the respective lengths of $r$ and $s$.
Alternatively, to avoid the expensive cost of determinization, one can instead create a nondeterministic finite automaton (NFA) for $r$~\cite{thompson} in $\bigo{\size{r}}$ and then perform matching with worst-case time complexity $\bigo{\size{r}\times\size{s}}$~\cite{regexp_simpl_fast,pike_sam}.

Modern regular expressions, which we call \textit{regexes} to differentiate them from traditional ones, have strayed far from their ancestors: they are significantly more complex and expressive, but much less well studied.
Their nontraditional features have allowed them to become one of the most pervasive embedded domain-specific languages in programming:
one study found them in more than 30\% of npm and PyPI packages~\cite{redos_impact};
another in 42\% of Python developments~\cite{survey_regex_python};
and 8 of the top 10 \textsc{Tiobe} languages support them natively~\cite{tiobe}.
One crucial extension is the introduction of capture groups, requiring regex engines to return not only whether there is a match, but also which part of the input string matches each sub-expression inside parentheses.
For instance, when matching \regex{(a*)b} on \str{caabd}, modern regex engines report a match on sub-string \str{aab}, and a sub-match on \str{aa} for the capture group \regex{(a*)}. 
Capture groups fundamentally transform the problem from one of language recognition to one of pattern matching: traditional regular expression matching checks whether an input string belongs to a formal language, whereas modern regex matching computes the positions of the sub-strings matched by the regex and each of its capture groups.
As such, regexes represent not a language, but a way to search and segment a piece of text.
This problem is often ambiguous, so modern regex languages define priority rules to deterministically disambiguate results when a regex can match a string in multiple ways.
For instance, when matching \regex{(a|a*)} on \str{aa}, most modern regex engines give priority to the left branch of the alternation and return a match only on the first letter (this differs from the longest-match semantics often found in lexers).

Other modern features include backreferences, lookarounds (lookaheads and lookbehinds, both positive and negative), anchors, character classes, and repetition.
Backreferences force a later part of a regex to match the same substring as an earlier group: for instance, \regex{(a*)b\textbackslash1} matches \str{aabaa} but not \str{aaba}, as \rgx{\textbackslash1} matches exactly the sub-string captured by the first group \regex{(a*)}.
Lookarounds condition a match without consuming the corresponding characters:
for instance, the regex \regex{(?<=£)1} matches the character \str{1} in \str{£1.2} and does not match anything in \str{v1.2} (the \str{£} that is checked for by the lookbehind is not included in the final match).
Anchors $~\hat{}~$, \rgx{\$}, and \rgx{\textbackslash{}b} match at the beginning and end of the string or at a word boundary,
character classes match ranges of characters (e.g., \rgx{[a-e]} matches any character between a and e),
and counted repetition repeat the same regex multiple times (\rgx{(a|b)\{4,8\}} matches any combinations of 4 to 8 \str{a} and \str{b}).
To illustrate the convenience of these features together, the regex \regex{(?<=PLDI)[0-9]\{2,4\}} matches the year of a reference to a PLDI paper (e.g., \str{2024} in \str{PLDI2024}).

These nontraditional features and their interplay are only partially understood.
It is well-known, for example, that backreferences make the matching problem NP-hard~\cite{nphard,aho90}.
At the other end of the spectrum, linear-time algorithms are in common use for features such as capture groups, anchors and character classes~\cite{re2,rust_regex}.
In-between is uncertainty: to the best of our knowledge, no such matching complexity result has been stated for, e.g., lookarounds with capture groups.
Accordingly, modern regex engines fall into two categories.
\textit{Backtracking} engines support all features (including backreferences) but suffer from algorithmic blowups --- even on simple patterns composed exclusively of traditional features for which linear-time algorithms are known.
This disastrous worst-case performance has serious security implications, ranging from partial service degradation to
complete outages of large websites~\cite{cloudflare, stackoverflow}.
A recent study~\cite{freezing_the_web} estimated that 12\% of JavaScript-based web servers are vulnerable to regex-based denial-of-service attacks, or ``ReDoS''.

These security concerns and complexity issues have led to revived interest in engines that match regexes with worst-case linear time guarantees, in exchange for a reduced feature set (no backreferences nor lookarounds~\cite{re2_unsupported}) \cite{dotnet_regex,rust_internals}.  Interest in these linear engines is causing a paradigm shift: an increasing number of platforms and languages are now either secure-by-default with linear engines (Rust~\cite{rust_regex} and Go, and any language linking to the popular RE2 library), or at least offer the possibility to switch to a linear engine for a subset of regexes (.NET~\cite{dotnet_pldi}).
Our work demonstrates that this same paradigm shift is applicable to JavaScript, and that the expressivity tradeoff is less dire than previously believed.

Unfortunately, even in the new, safer world, the word \emph{linear} still hides a lot of variability: some engines aim for linearity in just the input string length (e.g.\ .NET, so that performing matching in $\bigo{\size{r}^2\times\size{s}}$ is acceptable), whereas others attempt to guarantee linear execution in both \size{r} and \size{s} (e.g.\ Rust) and are therefore suitable for user-provided regexes or regexes derived from user input (for instance, Google Sheets evaluates user-provided regexes with RE2).  Our work achieves both regex- and string-linear performance.

To make matters worse, different regex languages make different semantic design choices when it comes to these nontraditional features~\cite{lingua_franca} and regexes written for a language are not always portable to another one: quantifiers (\rgx{*}, \rgx{+}, \rgx{?}) and capture groups have different semantics in JavaScript and Perl, most valid lookarounds in .NET and JavaScript are not valid in Python, Perl or Java (which prevent the use of quantifiers inside lookarounds as in \regex{(?<=ba*)}), etc.  The impact of these semantic design choices on matching complexity is not well understood.

This problem has several consequences.
From a programming language design point of view, the situation is dire.
If we ever want to design expressive but secure regex languages, we need a better understanding of the worst-case complexity of each nontraditional feature, \emph{under each semantic design choice}.
The situation is no better from a programming language implementation standpoint: even for features and semantics thought to be well understood in practice, our research shows that multiple deployed implementations make invalid algorithmic assumptions.
In particular, we found that some features commonly assumed not to be matchable in linear time could in fact be supported by linear engines; that algorithms assumed to be linear were in fact not; and that deployed algorithms assumed to be applicable to modern regex languages led to semantically incorrect results.
The following sections provide concrete examples of all three cases.

Our work focuses on the JavaScript regex language; its semantics is specified in the ECMAScript standard by means of a pseudocode backtracking algorithm~\cite{ecma_262}.
Javascript supports many nontraditional regex features, including capture groups, backreferences and lookarounds, yet its semantics also make some atypical choices (\S\ref{subsec:js_semantics}) that separate it from other languages and impact matching complexity (\S\ref{sec:linear_plus}).
All JavaScript regexes engines that we could find, except one, use a backtracking implementation strategy~\cite{quickjs, duktape, mujs, hermes, irregexp, webkit_yarr}.  The exception is V8 (the JavaScript implementation used in Google Chrome and Node.js), which has two engines: a full-featured backtracking engine called Irregexp~\cite{irregexp}, and a linear engine with support for almost all of JavaScript regex constructs except backreferences and lookarounds, which we call ``V8Linear''~\cite{non_backtrack_v8}, available through a command-line flag.\footnote{Mozilla recently abandoned the development of the regex engine used in Firefox and replaced it with Irregexp, owing to the complexity of JavaScript regex engines~\cite{spidermonkey_irregexp}.}
To the best of our knowledge, V8Linear is the only industrial-strength linear-time implementation of a significant subset of JavaScript regexes.

Our work answers the following research questions:
Which part of the JavaScript regex language can be matched with linear worst-case time complexity?
Can we get linearity in \emph{both} the size of the regex and the size of the string $\bigo{\size{r}\times\size{s}}$ ?
To what extent the semantic choices done in JavaScript have an impact on what features can be matched linearly?

Our contributions can be summarized as follows:
\begin{itemize}
\item We provide a \textbf{novel understanding of JavaScript regex semantic properties}, by identifying a large subset of JavaScript regexes that can be matched in linear time $\bigo{\size{r}\times\size{s}}$.
  We show that JavaScript's semantics for nullable quantifiers (\rgx{*,+,?}) is incompatible with traditional linear-matching techniques, and we present a way to adapt these techniques to achieve linearity.
  We further show that several JavaScript features (nullable plus and capture groups inside quantifiers) were not implemented linearly in the size of the regex in V8Linear and we present new algorithms to match most of these features linearly.
\item We introduce the \textbf{first nonbacktracking algorithms for matching lookarounds}, showing that traditional linear engines are too conservative in their feature set.
  We present an algorithm to match lookbehinds not containing capture groups in linear time that is independent of JavaScript's semantics and could be applied to other regex languages.
  We then show how to leverage a JavaScript-specific semantic property to match unrestricted lookarounds with capture groups in linear time, albeit with additional memory complexity.
  To the best of our knowledge, this is the first time that linear algorithms for lookarounds with capture groups have been presented and implemented.
\item We provide \textbf{practical implementations for all our algorithms}, showing that they are generally applicable and that their complexity can be validated through experiments.
  We implemented several of our algorithms in V8Linear,\footnote{Nonnullable plus \url{https://chromium-review.googlesource.com/c/v8/v8/+/4778506}, Nullable quantifiers \url{https://chromium-review.googlesource.com/c/v8/v8/+/4755530}, Lookbehinds \url{https://chromium-review.googlesource.com/c/v8/v8/+/5093860}.} with some having been merged and released in V8 already.
  We also present OCaml prototype implementations for all of the algorithms presented in this paper, together supporting a very large fragment of JavaScript regexes.
  This prototype provides an ideal playground for implementation experimentation and we use it to exhibit a novel trade-off between space complexity and time complexity for capture groups.
\end{itemize}

\def\streaming{\ensuremath{^{\hookrightarrow}}}
\def\plusmem{\ensuremath{^{\dag}}}
\begin{center}
\begin{tabular}{cccc}
  \textbf{JS Regex Feature} & \textbf{V8Linear (state of the art)} & \textbf{Our new complexity} & \\
  \hline
  Nullable quantifiers & incorrect & $\bigo{\size{r}\times\size{s}}$ & \S~\ref{sec:nullable}\\
  Capture Groups in Quantifiers & $\bigo{\size{r}^2\times\size{s}}$ & $\bigo{\size{r}\times\size{s}}$ & \S~\ref{sec:clocks}\\
  Nonnullable Plus & $\bigo{2^{\size{r}}\times\size{s}}$ & $\bigo{\size{r}\times\size{s}}$ & \S~\ref{sec:linear_plus}\\
  Nullable Greedy Plus & $\bigo{2^{\size{r}}\times\size{s}}$ & $\bigo{\size{r}\times\size{s}}$ & \S~\ref{sec:linear_plus}\\
  Captureless Lookbehinds & unsupported & $\bigo{\size{r}\times\size{s}}$ & \S~\ref{sec:memoryless}\\
  Unrestricted Lookarounds & unsupported & $\bigo{\size{r}\times\size{s}}\plusmem$ & \S~\ref{sec:lookarounds}\\
  \multicolumn{4}{c}{\footnotesize$\plusmem$: with an additional $\bigo{\size{r}\times\size{s}}$ space complexity}\\
\end{tabular}
\end{center}
This table summarizes most of our results.
Taken together, they show that \textbf{a large subset of JavaScript regexes can be matched in linear time}, for both the string and the regex size.

\section{Linear Regex Engines Background}
\label{sec:linear_engines}

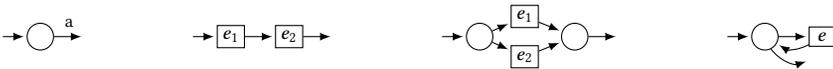
\begin{figure}[h]
  \centering
  \scriptsize
\begin{tikzpicture}[%
        every node/.style={circle,minimum size=10pt,minimum height=2pt, inner sep=2pt},
        shorten >=1pt,
        node distance=0.4cm, >=latex,
        initial text={}
      ]
  \node [initial] (charleft) [draw] {};
  \node [] (charright) [right=of charleft] {};
  \node [initial, rectangle] (conleft) [draw, right=of charright, xshift=1cm] {$e_1$};
  \node [rectangle] (conright) [draw, right=of conleft] {$e_2$};
  \node [] (conend) [right=of conright] {};
  \node [initial] (alt) [draw, right=of conend, xshift=1cm] {};
  \node [rectangle] (alt1) [draw, above right=of alt,yshift=-.3cm] {$e_1$};
  \node [rectangle] (alt2) [draw, below right=of alt,yshift=.3cm] {$e_2$};
  \node [] (altend) [] [draw, right=of alt, xshift=.5cm] {};
  \node [] (altnext) [] [right=of altend] {};
  \node [initial] (star) [draw, right=of altnext, xshift=1cm] {};
  \node [rectangle] (starbody) [draw, right=of star] {$e$};
  \node [] (starend) [below=of starbody,yshift=.4cm] {};
 
  \path [draw] (charleft) edge[->, above]  node {a} (charright);
  \path [draw] (conleft) edge[->]  node {} (conright);
  \path [draw] (conright) edge[->]  node {} (conend);
  \path [draw] (alt) edge[->]  node {} (alt1);
  \path [draw] (alt) edge[->]  node {} (alt2);
  \path [draw] (alt1) edge[->]  node {} (altend);
  \path [draw] (alt2) edge[->]  node {} (altend);
  \path [draw] (altend) edge[->]  node {} (altnext);
  \path [draw] (star) edge[->]  node {} (starbody);
  \path [draw] (starbody) edge[->, bend left]  node {} (star);
  \path [draw] (star) edge[->, bend right=35]  node {} (starend.west);
\end{tikzpicture}
\caption{Recursive Thompson NFA constructions for $a$, $e_1e_2$, $e_1|e_2$ and $e^*$.}
\label{fig:thompson_nfa}
  
\end{figure}

To help position our contributions, this section presents a brief overview of commonly used regex-matching algorithms.
Traditional regexes can be represented as nondeterministic finite automata (NFAs), with a mix of $\varepsilon$-transitions and transitions labeled with characters to read from the string.
A regex matches a string if there exists a path whose labeled transitions spell out the string in the corresponding NFA.
Figure~\ref{fig:thompson_nfa} summarizes the traditional Thompson construction~\cite{thompson} for traditional regexes (other constructions exist~\cite{glushkov}).
For a traditional regex of size $\size{r}$ (meaning that its textual representation uses $\size{r}$ characters), computing the Thompson NFA has $\bigo{\size{r}}$ time complexity and produces an NFA with $\bigo{\size{r}}$ states and transitions.
In the rest of this paper, we refer to a regex and its Thompson NFA interchangeably, so that the \textit{states} of a regex are those of its NFA.

\begin{wrapfigure}{L}{3cm}
  \centering
\begin{tikzpicture}[%
        every node/.style={circle,minimum size=10pt,minimum height=2pt, inner sep=2pt},
        shorten >=1pt,
        node distance=0.5cm, >=latex
      ]
  \node [] (s0) [] {};
  \node [rectangle] (s1) [draw, below=of s0] {\small\#1:entry};
  \node [] (s2) [draw, below left=of s1] {};
  \node [] (s3) [draw, below right=of s1] {};
  \node [rectangle] (s4) [draw, below=of s1, yshift=-.5cm] {\small\#1:exit};
  \node [accepting] (s6) [draw, below=of s4] {};
  \path [draw] (s0) edge[->, left]  node {} (s1);
  \path [draw] (s1) edge[->, dotted, left]  node {} (s2);
  \path [draw] (s1) edge[->, left]  node {} (s3);
  \path [draw] (s2) edge[->, left]  node {a} (s4);
  \path [draw] (s3) edge[->, right]  node {.} (s4);
  \path [draw] (s4) edge[->, left]  node {b} (s6);
\end{tikzpicture}
\scriptsize
\begin{lstlisting}[style=byt]
0: SetReg #1:entry
1: Fork 2 4
2: Consume 'a'
3: Jump 5
4: ConsumeAny
5: SetReg #1:exit
6: Consume 'b'
7: Accept
\end{lstlisting}  
\caption{Tagged NFA and its corresponding bytecode for \regex{(a|.)b}. The bytecode representation is explained later in Section~\ref{subsec:pikevm}.}
\label{fig:nfa_example}
\end{wrapfigure}
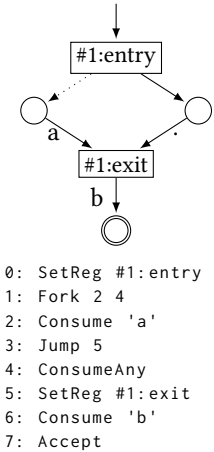

This traditional construction can be extended to handle capture groups.
First, nodes with multiple outgoing edges are augmented with a notion of edge priority, indicating which path should be considered first.
Second, effectful nodes are added to track the string positions at which each capture group is entered and exited.
An example of such a \emph{tagged} NFA~\cite{tagged_nfa} is shown in Figure~\ref{fig:nfa_example} (there and in the rest of this work dotted arrows indicate high priority edges).
A path through the \texttt{\#1:entry} node records at which string position it entered the first capture group.

Backtracking engines match regular expressions by enumerating all paths of the corresponding NFA in order of priority and returning the first accepting path.
This technique can naively be adapted to support all regex features (backreferences, lookarounds\dots), but it has worst-case exponential time complexity in the size of the string, and the worst case can happen even on regexes using only traditional regular expression features.
In the absence of backreferences (for which the matching problem is NP-hard for the size of both the input string and the regex~\cite{nphard,aho90}), and by excluding lookarounds and some other features, linear engines achieve linear-time complexity using the following insight:

\textbf{Uniform-futures property}:
Consider two path prefixes of an automaton $p_1$ and $p_2$. If they both reach the same regex state while having read the same prefix of the input string, then any extension of $p_1$ into a complete path is also a valid extension of $p_2$.
In the presence of capture groups, if $p_1$ has higher priority than $p_2$, then any extension of $p_1$ also has higher priority than $p_2$.
In other words, the future of a path prefix only depends on its current regex state and its current string position.

To illustrate this property, consider the regex \regex{(a+)*b} executed on the string \str{aaa}.  Let $p_1$ be the partial path matching the first two \str{a} with a single iteration of the star, and $p_2$ the partial path matching the first two \str{a} with \emph{two} iterations of the star.  Despite their diverging pasts, these paths have the same future: each can lead to a match if and only if the second path can also find one, and hence they do not need to be explored separately.
A typical backtracking engine will not perform such path merging: it will instead enumerate all possible decompositions of the input string into non-empty sub-strings.  Only at the end of each path will it notice that there is no \str{b} character to complete the match, leading to complexity exponential in the number of \str{a}.
All linear engines use the uniform-futures property to merge convergent paths. With backreferences, however, this uniform-futures property does not hold. The future of a path may depend on its prefix, because matching a backreference depends on how capture groups were previously matched. As a result, linear engines do not support backreferences. Until this paper, linear engines also excluded lookarounds. We show in Section~\ref{sec:lookarounds} that lookarounds can in fact be matched linearly.

\parbf{Bit-state backtracking (or memoized backtracking)}
The most straightforward linear-matching algorithm, bit-state backtracking, simply augments backtracking with a memoization table that records each pair of string position and regex state considered along the search.  This table is sufficient to leverage the uniform-futures property: if a path reaches a previously visited pair, it is simply discarded.
This approach preserves the usual benefits of backtracking engines (regexes that require little backtracking have excellent performance), but it comes at a significant additional memory cost $\bigo{\size{r}\times\size{s}}$.  Consequently, state-of-the-art engines use it only for small regex and strings~\cite{re2_wiki}.

\parbf{Thompson Simulation and PikeVM}
More memory-efficient is \emph{NFA simulation} (Thompson's algorithm~\cite{thompson}), which is used by most linear engines in the common case.  It explores the graph breadth-first, ensuring that paths that have consumed the same input prefix are considered at the same time, and leveraging the uniform-futures property to convergent paths.
NFA simulation extends to tagged NFAs (Figure~\ref{fig:nfa_example}) to support capture groups~\cite{regexp_vm_approach,pike_sam}, in which case the graph and its exploration algorithm are typically translated into a unified byte code, and executed by a so-called \emph{Pike VM} (Section~\ref{subsec:pikevm}).
This paper describes extensions to the NFA simulation and \emph{PikeVM} algorithms.

\parbf{Lazy DFA}
Finally, for regexes without capture groups, many engines perform \textit{lazy DFA simulation}, a variant of NFA simulation that performs better on average~\cite{re2_wiki} while maintaining worst-case linear complexity.  Lazy DFA performs similarly to NFA simulation but keeps visited states as \emph{sets} instead of lists (order is irrelevant when no groups are present).  The result can be viewed as interleaving traditional determinization and matching: it materializes the states that would have been visited by the traditional ahead-of-time DFA lazily, as it matches its input string (ahead-of-time construction has $\bigo{2^{\size{r}}}$ time complexity, whereas lazy DFA performs at most $\size{s}$ transitions, each of which have complexity $\size{r}$ to compute neighbor states).

\section{Technical Background}

We now move to our main focus: adapting NFA simulation to match JavaScript regular expressions.  This section first presents the JavaScript regex language and its semantics specificities, then
describes NFA simulation in detail, laying the groundwork for the presentation of our contributions in Section~\ref{sec:contributions}.

\subsection{JavaScript Regex Syntax}

\begin{wrapfigure}{L}{4.5cm}
    \small
    Regular Expressions:\\
    \begin{tabular}{l@{\hskip .1cm}r@{\hskip .1cm}l l}
\re & $::=$ & \chara & Character \\
&$\mid$& $.$ & Any Character\\
&$\mid$& $\eps$ & Empty\\
&$\mid$& $\re_1 ~ \re_2$ & Concatenation\\
&$\mid$& $\re_1 \trgx{|} \re_2$ & Union\\
&$\mid$& $\re ~ \quant$ & Quantifier \\
&$\mid$& $\trgx{(}\re\trgx{)}$ & Capture Group \\
&$\mid$& $\trgx{(?:}\re\trgx{)}$ & Noncapturing Group\\
&$\mid$& $\trgx{(}\look~\re\trgx{)}$ & Lookaround \\
    \end{tabular}\\
    Quantifiers:\\
    \begin{tabular}{l@{\hskip .1cm}r@{\hskip .1cm}l l}
\quant & $::=$& $\trgx{*}$ & Greedy Star\\
&$\mid$& $\trgx{*?}$ & Lazy Star\\
&$\mid$& $\trgx{+}$ & Greedy Plus\\
    \end{tabular}\\
    Lookarounds:\\
    \begin{tabular}{l@{\hskip .1cm}r@{\hskip .1cm}l l}
\look & $::=$& $\trgx{?=}$ & Positive Lookahead\\
&$\mid$& $\trgx{?!}$ & Negative Lookahead\\
&$\mid$& $\trgx{?<=}$ & Positive Lookbehind\\
&$\mid$& $\trgx{?<!}$ & Negative Lookbehind\\
    \end{tabular}
  \caption{Regex Syntax}
  \label{fig:syntax}
\end{wrapfigure}

This work considers the subset of JavaScript regexes depicted on Figure~\ref{fig:syntax}.
Greedy operators are also sometimes referred to as \textit{eager}, and lazy operators as \textit{nongreedy}.
All our algorithms also support additional features with linear-time guarantees, like anchors ($~\hat{}~$,\rgx{\$}), character classes (\rgx{[a-z]}, \rgx{\textbackslash d,\textbackslash w}\dots), or even arbitrary predicates that only consult the surrounding characters.
Counted repetitions are also supported by repeating the sub-expression (just like in state-of-the art linear engines).
This makes the regex and the matching time complexity grow with the counters used in counted repetitions.

\paragraph{Notations}
To make regexes more readable, we use the notation \rgx{\noncap{r}} for \textit{noncapturing groups}, instead of the usual \rgx{(?:r)} JavaScript syntax.
Such noncapturing groups are just annotations to make parsing unambiguous, but don't introduce any capture groups to extract.
Similarly, we use the character \rgx{\eps} to denote the empty regex, instead of not writing anything at all (\emph{e.g.} we write \regex{a|\eps} instead of \regex{a|}).
We sometimes annotate each regex capture group as follows: \regex{a(b)(c)} gets annotated to \regex{a(b)\grp{1}(c)\grp{2}}.
In JavaScript, each capture group must have a distinct identifier, even for named capture groups.
Similarly, we annotate each lookaround with an unique identifier.
Identifiers are integers given in a preorder AST traversal.
This ensures that if lookaround \rgx{l\lkr{i}} contains another lookaround \rgx{l\lkr{j}}, then $i<j$.
For instance, the regex \regex{(?=a(?<=a))a} gets annotated to \regex{(?=\lkr{1}a(?<=\lkr{2}a))a}.
We write \looknb{r} the total number of lookarounds in a regex $r$.
For instance, $\looknb{\regex{(?=a(?<=a))a}} = 2$.

\subsection{JavaScript Regex Semantics Peculiarities}
\label{subsec:js_semantics}

While most modern regex languages share a number of nontraditional features, they also present a number of differences~\cite{lingua_franca}.
These differences are scarcely documented, yet they can have a substantial impact on the semantics or time complexity of regex matching.
In this section, we focus on some subtle and notable properties of the JavaScript regex language, that may separate it from other modern regex languages.

\parbf{Priority}
The JavaScript backtracking semantics~\cite{ecma_262} explores paths in a given priority order.
In an alternation, the left branch has priority over the right one.
In a greedy quantifier, the priority is to iterate as much as possible, while a lazy quantifier tries to iterate as few times as possible.
JavaScript regexes are unanchored, meaning that matches can start anywhere in the input string.
The match that starts the earliest in the input string has the most priority.
In practice, engines add the prefix \rgx{.*?} to the regex they are executing to find that match.

\parbf{Capturing Lookarounds}\cite[22.2.2.4]{ecma_262}.
Lookarounds in JavaScript are more than assertions.
Lookahead and lookbehinds can define capture groups that the engine should return.
For instance, \regex{(?=(c)\grp{1})} on \str{c} returns that capture group \group{1} is set to \str{c}.
However, as in most modern regex languages with them, negative lookarounds cannot define any capture groups.
  

\parbf{Unbounded Lookarounds}\cite[22.2.1:Assertion]{ecma_262}
Some languages (\emph{e.g.} Perl, Python, Java) only allow fixed-width lookarounds. Meaning that regular expression patterns inside lookaheads and lookbehinds should not contain unbounded quantifiers like star and plus.
In JavaScript, there is no such restriction and \regex{(?=a*)} is a valid regex.
Fixed-width lookarounds are not much harder to implement than anchors, but unbounded ones are more expressive and complex.

\parbf{Capture Reset}\cite[22.2.2.3.1, step 4.]{ecma_262}
When entering a quantifier, the value of the capture groups defined inside that quantifier are reset to \undef.
For instance, matching \regex{((a)\grp{2}|(b)\grp{3})\grp{1}*} on string \str{ab} will return a match where group \group{2} is set to undefined.
On the first star iteration, \group{2} is set to the range \texttt{0-1}, matching the first character of the string.
When executing the second star iteration, \group{2} is reset to undefined, and group \group{3} is set to the range \texttt{1-2}, matching character \str{b}.
This property is specific to the JavaScript regex language.
We show that, while difficult to implement in linear time (see Section~\ref{sec:clocks}), this property helps implement other features in linear time (see Sections~\ref{subsec:cin_cdn} and~\ref{sec:lookarounds}).

\parbf{Nullable Quantifiers}\cite[22.2.2.3.1, step 2.b]{ecma_262}
Quantifiers can have both mandatory and optional iterations.
For instance, the plus has one mandatory iterations, followed by any number of optional ones.
The star has no mandatory iterations, but can have any number of of optional ones.
In JavaScript, optional repetitions of a quantifier cannot match the empty string.
This prevents the backtracking implementation from executing an infinite loop when matching a nullable star for instance.
This semantics for nullable quantifiers is specific to JavaScript.
Other regex languages typically have different semantics for nullable quantifiers that is discussed in Section~\ref{sec:nullable}.

\begin{figure}[b]
  \small
  \begin{minipage}{.4\textwidth}
  \textit{Usual NFA Simulation instructions:}\\
  \begin{tabular}{l@{\,}|@{\,}l}
    \bytecode{Consume}~\chara & Consumes character \chara.\\
    \bytecode{ConsumeAny} & Consumes any character.\\
    \bytecode{Jump}~\lbl & Jumps to label~\lbl.\\
    \bytecode{Fork}~\lbl$_1$~\lbl$_2$ & Creates a new thread. \\
    & \lbl$_1$ has higher priority.\\
    \bytecode{Accept} & A match is found.\\
    \bytecode{SetReg}~\reg & Writes current string \\
    & position to register \reg.\\
  \end{tabular}
  \end{minipage}%
  \begin{minipage}{.6\textwidth}
    \textit{New instructions introduced in this work:}\\
  \begin{minipage}{.5\textwidth}
    \begin{tabular}{l@{\,}|@{\,}l}
      \bytecode{BeginLoop} & \S~\ref{sec:nullable}\\
      \bytecode{EndLoop} & \S~\ref{sec:nullable}\\
      \bytecode{SetQuant}~$q$ & \S~\ref{sec:clocks}\\
      \bytecode{WriteOracle}~$l$ & \S~\ref{sec:lookarounds}\\
      \bytecode{CheckOracle}~$l$ & \S~\ref{sec:lookarounds}\\
      \bytecode{NegCheckOracle}~$l$ & \S~\ref{sec:lookarounds}\\
    \end{tabular}
  \end{minipage}%
  \begin{minipage}{.3\textwidth}
    \begin{tabular} {l@{\,}|@{\,}l}
      \bytecode{WriteLB}~$b$ & \S~\ref{sec:memoryless}\\
      \bytecode{CheckLB}~$b$ & \S~\ref{sec:memoryless}\\
      \bytecode{NegCheckLB}~$b$ & \S~\ref{sec:memoryless}\\
      \bytecode{SetNullPlus}~$q$ & \S~\ref{sec:linear_plus}\\
      \bytecode{CheckNull}~$q$ & \S~\ref{sec:linear_plus}\\
    \end{tabular}
  \end{minipage}
  \end{minipage}
  \captionof{figure}{NFA Simulation Bytecode Instructions}
  \label{fig:bytecode}
\end{figure}

\subsection{NFA Simulation Engines}
\label{subsec:pikevm}

\begin{minipage}[t]{.29\textwidth}
\begin{algorithm}[H]\scriptsize
\SetKwInOut{Input}{input}\SetKwInOut{Output}{output}
\SetKwFor{match}{match}{with}{end}
\SetKwFor{case}{case}{$\Rightarrow$}{}
\SetKw{hlet}{let}
\SetInd{0.1em}{0.75em}
\SetAlgoLined
\For{\texttt{i=0} \textbf{to} \texttt{str.length}}{
\While{\texttt{active}$\neq$\texttt{[]}}{
  \texttt{t = active.top()}\\
  \If{\texttt{processed[t.pc]}} {
    \texttt{active.pop()}\\
    \textbf{continue}
  }
  \texttt{processed[t.pc] = true}\\
  \match{\texttt{bytecode[t.pc]}}{
    \case{\bytecode{Consume c}}
         {\If{\texttt{c = str[i]}}{
             \texttt{t.pc = t.pc+1}\\
             \texttt{next.push(t)}}
           \texttt{active.pop()}}
    \case{\bytecode{Jump l1}}
         {\texttt{t.pc = l1}}
    \case{\bytecode{Fork l1 l2}}
         {\texttt{t.pc = l2}\\
           \texttt{t' = t.copy()}\\
           \texttt{t'.pc = l1}\\
           \texttt{active.push(t')}}
    \case{\bytecode{Accept}}
         {\texttt{bestmatch = t}\\
           \texttt{active = []}}
    \case{\bytecode{SetReg r}}
         {\texttt{t.regs[r] = i}\\
           \texttt{t.pc = t.pc+1}}    
  }
}
\texttt{active = next.reverse()}\\
\texttt{next = []}\\
\texttt{processed.fill(false)}
}
\Return \texttt{bestmatch}
\caption{\small Simulation}
\label{alg:pikevm}
\end{algorithm}
\end{minipage}\hfill%
\begin{minipage}[t]{.7\textwidth}
  The NFA simulation algorithm is a well-established way to avoid the exponential cost of determinization, used in V8Linear, RE2, Rust and Go.
  It is common to represent the tagged NFA with a bytecode.
This bytecode corresponds to an array of bytecode instructions depicted on Figure~\ref{fig:bytecode}, each associated with a label.

A simulation engine (as shown on Algorithm~\ref{alg:pikevm}) maintains a list of threads ordered by priority, \texttt{active} (a LIFO list).
Each thread contains a program counter \texttt{pc} and a set of registers \texttt{regs}.
Threads represent incomplete paths of the NFA synchronized at the same string position \texttt{i}.
Initially, \texttt{active} contains a single thread with \texttt{pc} 0.

The algorithm goes through the string one character at a time (the \textbf{for} loop).
To compute the \texttt{next} list of threads for the next string position, it follows transitions of the NFA, starting with the highest priority thread, using \texttt{active.top()}.
When a thread reaches a \bytecode{Consume} instruction that corresponds to the next character, it is pushed into \texttt{next}.
New threads may be created with the \bytecode{Fork} instruction.
If a thread reaches an \bytecode{Accept} instruction, it is stored as being the best possible match found so far.
Lower priority threads are discarded, but the algorithm keeps running with higher-priority threads in \texttt{next}.

The algorithm also maintains a \texttt{processed} array indicating which bytecode instruction has already been executed at this step.
Any thread that reaches a bytecode instruction already in that array (when \texttt{processed[t.pc]} is true) is discarded with \texttt{active.pop()} (using the uniform-futures property of section~\ref{sec:linear_engines}).
Thanks to this array, each bytecode instruction can be executed at most once for each string position.  
\end{minipage}

\paragraph{Example} Consider the regex \regex{(a|.)b} (whose bytecode is shown in Figure~\ref{fig:nfa_example}), with string \str{ab}.
Initially, \texttt{active} contains a single thread with \texttt{pc} 0.
After a first iteration for the first character, the \texttt{active} list contains a thread at \texttt{pc} 3, and another at \texttt{pc} 5.
During the second iteration, the second thread gets killed, as the first one executes the instruction at \texttt{pc} 5 first.
After reading the last character, a single thread reaches the \bytecode{Accept} instruction.
Using its register values, the capture group \group{1} is known to contain the sub-string delimited by indices 0 and 1, indicating the first letter \str{a}.
Note that on the string \str{ac}, the simulation engine would only check once if the thread in state \texttt{6} can accept the second character, and immediately conclude that there is no match.
A backtracking implementation would check it twice: one for each path that consumed the first character.

\paragraph{Complexity}
For a regex $r$, the size of the generated bytecode and its number of epsilon transitions grow linearly with the size of the regex $\bigo{\size{r}}$.
Then, at each string position, the simulation executes each bytecode instruction at most once, and follows each transition at most once.
It follows that, for a regex $r$ and a string $s$, this executes in the worst case $\bigo{\size{r}\times\size{s}}$ bytecode instructions.
All of these instructions, except \bytecode{Fork}, are trivially implemented in $\bigo{1}$ time complexity.
The case of \bytecode{Fork} and the space complexity are discussed later in Section~\ref{sec:tradeoff}.

\section{Matching JavaScript Regexes with Linear-Time Guarantees}
\label{sec:contributions}

In this section, we present six different algorithms to match different features of the JavaScript regex language with linear time guarantees.
We first show in Section~\ref{sec:nullable} that the Nullable Quantifier semantic property of JavaScript is incompatible with a traditional NFA simulation engine, but present an extension that solves the issue while retaining linear complexity.
We then present in Section~\ref{sec:clocks} an algorithm to implement the Capture Reset semantic property in linear time, while this previously introduced regex-quadratic time complexity in other implementations.
Next, we introduce previously unsupported JavaScript features in a linear engine.
Section~\ref{sec:lookarounds} presents an algorithm for unrestricted lookarounds, while Section~\ref{sec:memoryless} presents a streaming algorithm for lookbehinds without capture groups.
We then present algorithms for linear matching of any nonnullable or greedy JavaScript plus (Section~\ref{sec:linear_plus}).
Finally, we exhibit a novel space and time complexity tradeoff when implementing capture group registers with NFA simulation (Section~\ref{sec:tradeoff}).
Our algorithms are composable (Section~\ref{sec:composition}).
Taken together, we show that a vast majority of JavaScript regexes without backreferences can be matched in $\bigo{\size{r}\times\size{s}}$ worst-case time complexity. 

\subsection{Matching JavaScript Nullable Quantifiers in a Linear Engine}
\label{sec:nullable}

\begin{wrapfigure}{l}{3cm}
  \scriptsize
\begin{tikzpicture}[%
        every node/.style={circle,minimum size=10pt,minimum height=2pt, inner sep=2pt},
        shorten >=1pt,
        node distance=0.5cm, >=latex
  ]
  \node [] (entry) [inner sep=0pt, minimum size=0pt, minimum height=0pt] {};
  \node [] (fork0) [draw, below=of entry, yshift=.2cm] {$s_0$};
  \node [] (fork1) [draw, below=of fork0] {$s_1$};
  \node [accepting] (accept) [draw, right=of fork0] {$s_6$};
  \node [] (alta) [draw, below left=of fork1] {$s_2$};
  \node [] (smid) [draw, below right=of alta] {$s_3$};
  \node [] (altb) [draw, below right=of smid] {$s_4$};
  \node [] (endloop) [draw, below left=of altb] {$s_5$};
  \path [draw] (entry) edge[->] node {} (fork0);
  \path [draw] (fork0) edge[->, dotted]  node {} (fork1);
  \path [draw] (fork0) edge[->] node {} (accept);
  \path [draw] (fork1) edge[->, dotted] node {} (alta);
  \path [draw] (fork1) edge[->] node {} (smid);
  \path [draw] (alta) edge[->, left]  node {a} (smid);
  \path [draw] (smid) edge[->] node {} (altb);
  \path [draw] (smid) edge[->, dotted] node {} (endloop);
  \path [draw] (altb) edge[->, right] node {b} (endloop);
  \path [draw] (endloop) edge[->, bend left=110] node {} (fork0);
\end{tikzpicture}
\caption{Usual NFA for \regex{\noncap{\noncap{a|\eps}\noncap{\eps|b}}*}}
\label{fig:nfa_bug}
\end{wrapfigure}

Among all regex languages, JavaScript has a unique semantics when it comes to matching nullable quantifiers (\rgx{*} or \rgx{+} for instance).
Surprisingly, the techniques and the uniform-futures property presented in section~\ref{sec:linear_engines} do not obey these semantics.
As a result, the V8Linear engine (using NFA simulation) was incorrect and would sometimes return a different result than specified.
However, we present a way to adapt these algorithms without changing their asymptotic complexity.

The NFA simulation algorithm cannot visit the same regex state twice without consuming any character in the string (see the \texttt{processed} array of Algorithm~\ref{alg:pikevm}).
In JavaScript however, it is possible to visit the same regex state twice without consuming, as long as this state does not mark the beginning of a quantifier.
Consider for instance matching the regex \regex{\noncap{\noncap{a|\eps}\noncap{\eps|b}}*} on the string \str{ab} (its NFA is represented on Figure~\ref{fig:nfa_bug}).
The top priority result of an usual NFA simulation is to match only \str{a} in a single star iteration.
Doing a second iteration of the star is invalid, since it would visit the same regex state ($s_3$) without having consumed anything from the string yet.

\begin{wrapfigure}{L}{5cm}
  \scriptsize
\begin{tikzpicture}[%
        every node/.style={circle,minimum size=10pt,minimum height=2pt, inner sep=2pt},
        shorten >=1pt,
        node distance=0.45cm, >=latex
  ]
  \node [] (entry) [inner sep=0pt, minimum size=0pt, minimum height=0pt] {};
  \node [] (fork0) [draw, below=of entry,yshift=.2cm] {$s_0$};
  \node [] (begin) [draw, rectangle, below=of fork0] {\bytecode{BeginLoop}};
  \node [] (fork1) [draw, below=of begin] {$s_1$};
  \node [accepting] (accept) [draw, right=of fork0] {$s_6$};
  \node [] (alta) [draw, below left=of fork1] {$s_2$};
  \node [] (smid) [draw, below right=of alta] {$s_3$};
  \node [] (altb) [draw, below right=of smid] {$s_4$};
  \node [] (endloop) [draw, rectangle, below=of smid, yshift=-.5cm] {\bytecode{EndLoop}};
  
  \node [] (fork0_) [draw, right=of fork0, xshift=1.5cm] {$t_0$};
  \node [] (begin_) [draw, rectangle, below=of fork0_] {\bytecode{BeginLoop}};
  \node [] (fork1_) [draw, below=of begin_] {$t_1$};
  \node [accepting] (accept_) [draw, right=of fork0_] {$t_6$};
  \node [] (alta_) [draw, below left=of fork1_] {$t_2$};
  \node [] (smid_) [draw, below right=of alta_] {$t_3$};
  \node [] (altb_) [draw, below right=of smid_] {$t_4$};
  \node [] (endloop_) [draw, rectangle, below=of smid_, yshift=-.5cm] {\bytecode{EndLoop}};

  \path [draw] (entry) edge[->] node {} (fork0);
  \path [draw] (fork0) edge[->, dotted]  node {} (begin);
  \path [draw] (begin) edge[->] node {} (fork1_);
  \path [draw] (fork0) edge[->] node {} (accept);
  \path [draw] (fork1) edge[->, dotted] node {} (alta);
  \path [draw] (fork1) edge[->] node {} (smid);
  \path [draw] (alta) edge[->, left]  node {a} (smid);
  \path [draw] (smid) edge[->] node {} (altb);
  \path [draw] (smid) edge[->, dotted] node {} (endloop);
  \path [draw] (altb) edge[->, right] node {b} (endloop);
  \path [draw] (endloop.west) edge[->, bend left=40] node {} (fork0);
  
  \path [draw] (fork0_) edge[->, dotted]  node {} (begin_);
  \path [draw] (begin_) edge[->] node {} (fork1_);
  \path [draw] (fork0_) edge[->] node {} (accept_);
  \path [draw] (fork1_) edge[->, dotted] node {} (alta_);
  \path [draw] (fork1_) edge[->] node {} (smid_);
  \path [draw] (alta_) edge[->, above]  node {a} (smid);
  \path [draw] (smid_) edge[->] node {} (altb_);
  \path [draw] (smid_) edge[->, dotted] node {} (endloop_);
  \path [draw] (altb_) edge[->, above] node {b} (endloop);
\end{tikzpicture}
\caption{Fixed NFA for \rgx{\noncap{\noncap{a|\eps}\noncap{\eps|b}}*}}
\label{fig:nfa_double}
\end{wrapfigure}

In JavaScript, the correct result is to match the entire string \str{ab} in two iterations of the star.
The first iteration matches \str{a} in the first alternation and the empty string in the second alternation.
The second iteration matches $\epsilon$ then \str{b}.
Both of these iterations of the star have matched a nonempty part of the string.
In JavaScript, the future of a path prefix depends not only on the current string position and the current regex state, but also on what has been consumed in the current quantifier.
Going from $s_3$ to $s_3$ without consuming characters in Figure~\ref{fig:nfa_bug} is only possible when the current execution has consumed \str{a} in $s_2$ before reaching $s_3$.

To adapt the NFA simulation algorithm to JavaScript semantics, we observe that, given a regex state and a string position, there are at most two possible behaviors depending on whether the current path is allowed to exit a quantifier without consuming a character.
To materialize these two behaviors, we duplicate the states of the regex, as shown on Figure~\ref{fig:nfa_double}.
The sub-automaton on the left of the Figure contains states $s_i$ that are allowed to exit any quantifier without consuming a character, because they already consumed a character in all of their parent quantifiers.
States $t_i$ of the right sub-automaton however are states that should not be allowed to exit their current innermost quantifier without consuming.
A thread that just began a quantifier iteration should not be able to exit this iteration just yet.
Consequently, at the beginning of each quantifier, we insert a new \bytecode{BeginLoop} bytecode instruction, which switches the execution to the automaton on the right.
When consuming a character, transitions always point back to the automaton on the left.
Even if a thread was inside several nested quantifiers, all of them just consumed a character and are then allowed to conclude their current iteration.
At the end of each quantifier, we insert a new \bytecode{EndLoop} instruction.
In the left automaton, this behaves like a \bytecode{Jump}, but on the right automaton this is a blocking state.

\setlength{\algomargin}{0em}%
\setlength{\interspacetitleruled}{0pt}%
\setlength{\algotitleheightrule}{0pt}%

\begin{wrapfigure}{L}{2.5cm}
  \begin{minipage}{2.5cm}
\begin{algorithm}[H]\scriptsize
\SetKwFor{case}{case}{$\Rightarrow$}{}
\SetKw{hlet}{let}
\SetInd{0.1em}{0.75em}
\SetAlgoLined
\case{\bytecode{Consume c}}
     {\If{\texttt{c=str[i]}}{
         \texttt{t.pc = t.pc+1}\\
         \texttt{t.left = true}\\
         \texttt{next.push(t)}}
       \texttt{active.pop()}}
\case{\bytecode{BeginLoop}}
     {\texttt{t.left = false}\\
       \texttt{t.pc = t.pc+1}}
\case{\bytecode{EndLoop}}
     {\eIf{\texttt{t.left}}{
         \texttt{t.pc = t.pc+1}
       }
       {\texttt{active.pop()}}}
\end{algorithm}
  \end{minipage}
\caption{Executing \bytecode{BeginLoop}/\bytecode{EndLoop}}
\label{fig:beginend}
\end{wrapfigure}

\paragraph{Correctness}
Regex quantifiers are well-parenthesized, meaning that it's never possible for a thread to be allowed to exit its innermost quantifier but not an outermost one.
As a result, no matter the number of nested quantifiers, two copies of the original automaton are enough to capture all the behaviors of the JavaScript quantifier semantics.

We reported this semantic mismatch, then implemented and merged our solution in V8Linear.
In practice in an NFA simulation implementation, one does not need to actually duplicate the bytecode.
Instead, threads are augmented with a boolean indicating in which automaton they currently belong.
\bytecode{BeginLoop}, \bytecode{EndLoop} and \bytecode{Consume} instructions each modify or read this boolean.
An extension of Algorithm~\ref{alg:pikevm} is shown on Figure~\ref{fig:beginend}, where this boolean is called \bytecode{left}.
When the regex inside a quantifier is not nullable (cannot match the empty string), there is no need to insert \bytecode{BeginLoop} and \bytecode{EndLoop} instructions.
A simple analysis to determine if a regex is non nullable is shown later in Section~\ref{sec:linear_plus}.
We illustrated our solution on NFA simulation, but the same insight can be used for a bit-state backtracker or a Lazy DFA matcher.
This also generalizes to counted repetition, where V8Linear used to return incorrect results.
For instance, in \regex{\noncap{\noncap{a|\eps}\noncap{\eps|b}}\{0,7\}}, the optional repetitions are not allowed to match the empty string.
It suffices to wrap the bytecode of each optional repetition with \bytecode{BeginLoop} and \bytecode{EndLoop} instructions.

\subsection{Linear Matching of the Capture Reset Property}
\label{sec:clocks}

The Capture Reset property introduced in section~\ref{subsec:js_semantics} is unique to the JavaScript regex language.
In this section we show how the solution used in V8Linear has quadratic complexity in the size of the regex, and we present a new linear algorithm for the Capture Reset property.

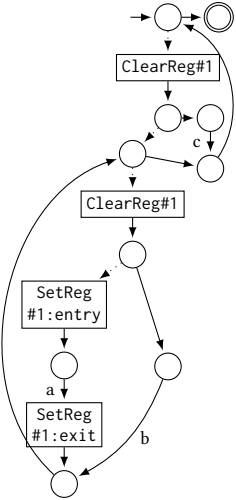
\begin{wrapfigure}{L}{4cm}
  \scriptsize
\begin{tikzpicture}[%
        every node/.style={circle,minimum size=10pt,minimum height=2pt, inner sep=2pt,align=center},
        shorten >=1pt,
        node distance=0.3cm, >=latex,
        initial text={}
      ]
  \node [] (entry') [draw,xshift=3cm] {};
  \node [rectangle] (clear') [draw, below=of entry'] {\bytecode{ClearReg\#1}};
  \node [] (fork') [draw, below=of clear'] {};
  \node [rectangle] (set1') [draw, below left=of fork'] {\bytecode{SetReg}\\\bytecode{\#1:entry}};
  \node [] (consa') [draw, below=of set1'] {};
  \node [rectangle] (set2') [draw, below=of consa'] {\bytecode{SetReg}\\\bytecode{\#1:exit}};
  \node [] (consb') [draw, below right=of fork', yshift=-1cm] {};
  \node [] (join') [draw, below=of set2'] {};
  \node [] (outfork) [draw, above right=of entry'] {};
  \node [rectangle] (outclean) [draw, above=of outfork] {\bytecode{ClearReg\#1}};
  \node [initial] (outentry) [draw, above=of outclean] {};
  \node [accepting] (outexit) [draw, right=of outentry] {};
  \node [] (consc) [draw, right=of outfork, xshift=-.1cm] {};
  \node [] (outjoin) [draw, below=of consc] {};
  
  \path [draw] (entry') edge[->, dotted]  node {} (clear');
  \path [draw] (clear') edge[->] node {} (fork');
  \path [draw] (fork') edge[->,dotted] node {} (set1');
  \path [draw] (fork') edge[->] node {} (consb');
  \path [draw] (set1') edge[->] node {} (consa');
  \path [draw] (consa') edge[->,left] node {a} (set2');
  \path [draw] (consb') edge[->,right,bend left=20] node {b} (join');
  \path [draw] (set2') edge[->] node {} (join');
  \path [draw] (join') edge[->,bend left=60] node {} (entry');
  \path [draw] (entry') edge[->] node {} (outjoin);
  \path [draw] (outentry) edge[->] node {} (outexit);
  \path [draw] (outentry) edge[->,dotted] node {} (outclean);
  \path [draw] (outclean) edge[->] node {} (outfork);
  \path [draw] (outfork) edge[->,dotted] node {} (entry');
  \path [draw] (outfork) edge[->] node {} (consc);
  \path [draw] (consc) edge[->,left] node {c} (outjoin);
  \path [draw] (outjoin) edge[->, bend right=50] node {} (outentry);
  
\end{tikzpicture}
\caption{\regex{\noncap{\noncap{(a)|b}*|c}*} NFA with \bytecode{ClearReg} instructions.}
\label{fig:clear_nfa}
\end{wrapfigure}

\paragraph{The previous quadratic algorithm}
The intuitive solution used in V8Linear defines a new bytecode instruction, \bytecode{ClearReg}~\reg, which clears the value of a capture register, setting the corresponding group to \undef.
Such instructions are inserted at the beginning of each quantifier, for each capture group defined inside that quantifier.
For instance, Figure~\ref{fig:clear_nfa} shows the bytecode instructions generated for \regex{\noncap{\noncap{(a)|b}*|c}*}.
Here we omit the \bytecode{BeginLoop} and \bytecode{EndLoop} instructions of section~\ref{sec:nullable} to explain both properties independently.
When a capture group is defined inside several quantifiers, one needs to clear its capture registers at the beginning of each of these quantifiers.
For instance, when matching \regex{\noncap{\noncap{(a)|b}*|c}*} on \str{ac}, the first \bytecode{ClearReg} instruction is needed to clear the capture group as we enter the outer star a second time.
When matching the same regex on \str{ab}, the second \bytecode{ClearReg} instruction is needed to clear the value of the group as we enter the inner star a second time.
As a result, each capture group might need as many bytecode instructions as the number of quantifiers above in the regex AST.
Consider the following family of regexes: $r_0=\regex{.}$ and $r_{n+1}=\regex{($r_n$)*}$.
While its size $\size{r_n}$ grows linearly in $\bigo{n}$, its bytecode size grows quadratically in $n$ (with $\bigo{n^2}$ complexity).

\paragraph{Our linear algorithm}
We now present our solution to implement the Capture Reset property linearly in the size of the regex with an NFA simulation.
In essence, it consists in not clearing the capture group values during the execution of the bytecode for all threads, but instead after a match is found, only for the winning thread.
Consider the regex \regex{\noncap{\noncap{(a)|b}*|c}*} again.
If a match is found with some value for the capture group, there are only two reasons for which we would like to clear its value:
either the inner star or the outer star were entered again after the capture group was set.
If we know when was the last time each quantifier and each group were entered we can filter accordingly, keeping the capture value only if it was defined later than the times both quantifiers were entered.
To define this notion of time, we extend the NFA simulation engine with a global \textit{clock}, an integer value starting at 0 and increasing each time the simulation executes any bytecode instruction.
We show on Figure~\ref{fig:clk} how to extend Algorithm~\ref{alg:pikevm}, where \bytecode{clk} is that clock.
We also extend the threads memory with some new registers containing clock values.
One new register for each group (in \bytecode{gclocks}), and one for each quantifier (in \bytecode{qclocks}).
Whenever the NFA simulation executes a \bytecode{SetReg} instruction corresponding to the entry of a capture group, it records both the current string position and the current clock value.
We also add a new \bytecode{SetQuant}~$q$ instruction, inserted at the beginning of each quantifier body, that records the clock value for quantifier $q$.
As a result, the size of the bytecode corresponding to a quantifier is now constant and does not depend on how many capture groups are defined inside.
This requires more registers for each thread, but as both the number of groups and the number of quantifiers are bounded by $\size{r}$, this memory still grows linearly with the size of the regex.

\begin{wrapfigure}{L}{3.8cm}
  \small
\begin{tikzpicture}[%
        every node/.style={circle,minimum size=10pt,minimum height=2pt, inner sep=2pt,align=center},
        node distance=0.3cm, >=latex,
        initial text={}
      ]
  \node [] (outer) [] {\rgx{*}:20};
  \node [] (left) [below left=of outer,xshift=-.3cm] {\rgx{*}:5};
  \node [] (right) [below right=of outer,xshift=.3cm] {\rgx{*}:35};
  \node [] (a) [below=of left, yshift=0.1cm] {\rgx{(a)}:6};
  \node [] (b) [below left=of right] {\rgx{(b)}:29};
  \node [] (c) [below right=of right] {\rgx{(c)}:37};
  \path [draw] (outer) edge[-] node {} (left);
  \path [draw] (outer) edge[-] node {} (right);
  \path [draw] (left) edge[-] node {} (a);
  \path [draw] (right) edge[-] node {} (b);
  \path [draw] (right) edge[-] node {} (c);
\end{tikzpicture}
\caption{Simplified AST for \regex{\noncap{(a)*|\noncap{(b)|(c)}*}*} with clock values of the winning thread for \str{abc}.}
\label{fig:simpl_ast}
\end{wrapfigure}
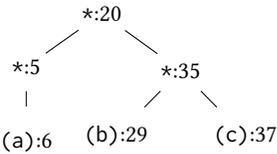

Filtering the capture groups after a match is found can be done with time complexity $\bigo{\size{r}}$, with an AST traversal of $r$.
For instance, consider the regex \regex{\noncap{(a)*|\noncap{(b)|(c)}*}*} and the string \str{abc}.
Figure~\ref{fig:simpl_ast} represents a simplified AST of the regex, where we wrote the final clock values of each group and quantifier in the winning thread.
We start with the root of the AST, and compare its value with both its children.
Because the inner left star has a smaller clock value than the outer star, all capture groups inside are cleared.
We then recursively consider the inner right star and compare its last clock value to its children.
One of the groups, \rgx{(b)}, has a smaller clock value, meaning that it was not defined in the last iteration of its parent star and is cleared.
Finally, we keep the value of the last group \rgx{(c)}.

\begin{wrapfigure}{L}{2.6cm}
  \begin{minipage}{2.6cm}
\begin{algorithm}[H]\scriptsize
\SetKwFor{case}{case}{$\Rightarrow$}{}
\SetKw{hlet}{let}
\SetInd{0.1em}{0.75em}
\SetAlgoLined
\case{\bytecode{SetReg r}}
     {\texttt{t.regs[r] = i}\\
       \texttt{t.gclocks[r] = clk}\\
       \texttt{t.pc = t.pc+1}}    
\case{\bytecode{SetQuant}}
     {\texttt{t.qclocks[r] = clk}\\
       \texttt{t.pc = t.pc+1}}
\end{algorithm}
  \end{minipage}
\caption{Updating clocks}
\label{fig:clk}
\end{wrapfigure}

\paragraph{Correctness}
Our new NFA construction does not change the paths explored by the threads, it only replaces some \bytecode{ClearReg} instructions with a single \bytecode{SetQuant} instructions.
The correctness of the new filtering algorithm relies on the following informal invariant:
at any moment during the NFA simulation, for any thread $t$, any group $g$ inside a quantifier $q$, the clock value of $g$ in $t$ is strictly greater than the clock value of $q$ if and only if the capture value of $g$ was defined in the last iteration of $q$ in $t$.
This invariant holds even when $g$ is inside several nested quantifiers.

The \bytecode{SetQuant} instructions can be omitted for quantifiers with no capture group inside, ensuring that our new bytecode generation is always smaller than the previous solution with \bytecode{ClearReg} instructions.
While we present this solution on an NFA simulation engine, the same idea could be applied to a backtracking or a bit-state backtracking implementation, as both may also spend a quadratic amount of time clearing capture registers if they follow the implementation described in the JavaScript semantics.
Note however that in a backtracking implementation supporting backreferences, groups that may be backreferenced need to be cleared dynamically, so that the backreference matches the correct value (in JavaScript, a backreference to an \undef~ group matches the empty string).

\subsection{Unrestricted JavaScript Lookarounds in Linear Time}
\label{sec:lookarounds}

We now present an algorithm to match all JavaScript lookarounds (both lookaheads and lookbehinds, both positive and negative) in linear time.
JavaScript lookarounds have two roles: filtering matches and defining capture groups.
Our algorithm handles these roles separately in different phases.
First, lookarounds act as assertions filtering matches.
For this aspect, we show that we can precompute an \textit{oracle} boolean truth table, indicating each string position at which each lookaround holds.
We show that constructing this oracle can be done in $\bigo{\size{r}\times\size{s}}$ complexity.
We then match the main expression, simply consulting the oracle when a path reaches a lookaround.
Finally, we reconstruct missing capture groups defined inside lookarounds in a third phase.
Thanks to the Capture Reset property of JavaScript, we show that we can also do that in $\bigo{\size{r}\times\size{s}}$ complexity.
An example of executing all phases of our algorithm is provided in the supplemental material.
This algorithm comes with an additional space complexity of $\bigo{\looknb{r}\times\size{s}}$ for the oracle where $\looknb{r}\le\size{r}$ is the number of lookarounds in $r$.
We will show in Section~\ref{sec:memoryless} how to avoid this space complexity in the particular case of captureless lookbehinds.

\subsubsection{First Phase: Building the Oracle}
We define the \textit{intrinsic size} of a regex $r$, noted \psize{r}, the size of its textual represenation without descending recursively inside lookarounds.
We get the following \textbf{intrinsic equality}, expressing that the full size of a regex is equal to the sum of its intrinsic size and the intrinsic sizes of all its lookarounds:$\forall r, \size{r} = \psize{r} + \sum_{i=1}^{\looknb{r}}\psize{r\lkr{i}}$.

To build each row of the oracle, we present a modification of the NFA simulation algorithm that allows to find in $\bigo{\psize{r\lkr{i}}\times\size{s}}$ all the places in $s$ where $r\lkr{i}$ matches.
In essence, we match \rgx{r\lkr{i}} \textit{in reverse} and modify the \bytecode{Accept} instruction.

\par\textbf{Observation 1:}
Replacing the final \bytecode{Accept} instruction of an NFA by an instruction that writes the current string position and does not discard lower priority threads allows the NFA simulation algorithm to find all positions at which a match can \textit{end}.
We consequently define the instruction \bytecode{WriteOracle i} which writes to the oracle.

\par\textbf{Observation 2:}
Define the reverse of \rgx{r\lkr{i}}, $\mathit{rev}(\rgx{r\lkr{i}})$ by recursively inverting the two subexpressions of each concatenation in \rgx{r\lkr{i}}.
Matching \regex{.*?$\mathit{rev(\rgx{r\lkr{i}})}$} on the reverse of a string $s$ can find all positions where a match of $\rgx{r\lkr{i}}$ in $s$ could begin.
Consequently, we extend the NFA simulation algorithm by allowing it to read the input string backward, starting with the last character of the string, then moving toward its beginning.
We use the standard forward direction for lookbehinds, and the backward direction for lookaheads that we also reverse.

Using these observations, we can compute each string positions at which each lookaround hold (\emph{i.e.} the positions at which a match can begin).
We construct the oracle one row at a time, starting with lookarounds with the highest indices (the deepest in the AST).
For nested lookarounds, the row of the inner lookarounds gets computed before its parent.
The parent can then replace its inner lookaround $\rgx{r\lkr{i}}$ by a single \bytecode{CheckOracle i} instruction, which checks the oracle table at the current string position.
Doing so, we build the entire row $i$ of the oracle table with a single match of \rgx{r\lkr{i}} on the string.\footnote{As an optimization, note that capture groups do not matter in that step, and we can freely remove them from each lookaround subexpression. We can even use a Lazy DFA matcher instead of the NFA simulation for faster matching.}
The complexity of building the entire table is then $\bigo{\sum_{i=1}^{\looknb{r}}\psize{r\lkr{i}}\times\size{s}}$, bounded by $\bigo{\size{r}\times\size{s}}$ using the intrinsic equality.




\subsubsection{Second Phase: Matching the main expression}

Once we know the positions at which each lookaround holds, we can simply run the main expression in a forward direction.
Whenever a thread encounters a lookaround, it simply accesses the oracle with a \bytecode{CheckOracle} instruction.
When the NFA simulation executes this instruction, it also records in each thread the string position at which each lookaround was last used.
This means defining new thread registers, one for each of $\bigo{\size{r}}$ many lookarounds, just as the ones used for capture groups (no change to the asymptotic complexity).
This phase has time complexity $\bigo{\psize{r}\times\size{s}}$.
When a match is found, this phase returns both the register values for all capture groups defined inside the main expression,
and the last string position each lookaround was used.
We reconstruct the values of capture groups defined inside positive lookarounds in the third phase.

\subsubsection{Third Phase: Reconstructing Missing Capture Groups}

Finally, we run an NFA simulation for each outermost lookaround that was used to produce the main match.
This time, this simulation is run forward for lookaheads and backward for lookbehinds (that we reverse), to comply with the capture semantics inside lookarounds.
We start the simulation exactly at the input string position where the lookaround was used.
If lookarounds are inside quantifiers, they may have been used several times in a match.
However, the Capture Reset property ensures that only its last visit can define the capture groups inside.
As we run the simulation for an outer lookaround, it may require going through an inner lookaround.
In that case, we check the oracle with a single \bytecode{CheckOracle} instruction, mark the inner lookaround and execute it later.
Winning threads of these new simulations define the missing capture group values.
In the worst-case, each lookaround needs to be executed once in this phase, which has time complexity
$\bigo{\sum_{i=1}^{\looknb{r}}\psize{r\lkr{i}}\times\size{s}}$, or $\bigo{\size{r}\times\size{s}}$ using the intrinsic equality.

\begin{wrapfigure}{L}{2.6cm}
  \vspace{-0.3cm}
  \begin{minipage}{2.6cm}
\begin{algorithm}[H]\scriptsize
\SetKwFor{case}{case}{$\Rightarrow$}{}
\SetKw{hlet}{let}
\SetInd{0.1em}{0.75em}
\SetAlgoLined
\case{\bytecode{WriteOracle l}}
     {\texttt{oracle[l][i] = true}\\
       \texttt{active.pop()}}    
\case{\bytecode{CheckOracle l}}
     {\eIf{oracle[l][i]}
       {t.pc = t.pc + 1}
       {active.pop()}}
\end{algorithm}
  \end{minipage}
\caption{Oracle}
\label{fig:writecheck}
\end{wrapfigure}

Figure~\ref{fig:writecheck} shows how to augment Algorithm~\ref{alg:pikevm} to support the new instructions \bytecode{WriteOracle} and \bytecode{CheckOracle}. Note that, unlike an \bytecode{Accept}, \bytecode{WriteOracle} does not kill lower priority treads but only the current one (and writes to the oracle), in accordance with Observation 1.

\paragraph{Correctness}
Observation 1 follows from the uniform-futures property, and Observation 2 can be proved by induction on the regex.
The correctness of the first two phases directly follows from Observations 1 and 2 and the correctness of the standard NFA simulation algorithm.
The third phase follows from the Capture Reset property: capture groups inside lookarounds can only be defined by the last iteration of this lookaround in the winning thread.

\subsection{Matching Captureless Lookbehinds with an NFA Simulation}
\label{sec:memoryless}

The previous algorithm of Section~\ref{sec:lookarounds} requires precomputing the entire oracle table.
For lookbehinds without capture groups inside, this is not needed.
In this section, we present a novel separate \textit{streaming} algorithm to extend the NFA simulation algorithm to handle captureless lookbehinds.
In particular, all negative lookbehinds are supported, as they cannot define capture groups (see Section~\ref{subsec:js_semantics}).
Compared to the previous section, this algorithm needs no additional space complexity.
It does not use the Capture Reset property and is thus applicable to any regex language, not just JavaScript.
This algorithm supports nested lookbehinds and capture groups outside of lookbehinds.

\begin{wrapfigure}{L}{2.4cm}
  \begin{minipage}{2.4cm}
\begin{algorithm}[H]\scriptsize
\SetKwFor{case}{case}{$\Rightarrow$}{}
\SetKw{hlet}{let}
\SetInd{0.1em}{0.75em}
\SetAlgoLined
\case{\bytecode{WriteLB b}}
     {\texttt{LBtable[b] = true}\\
       \texttt{active.pop()}}    
\case{\bytecode{CheckLB b}}
     {\eIf{LBtable[b]}
       {t.pc = t.pc + 1}
       {active.pop()}}
\end{algorithm}
  \end{minipage}
\caption{\bytecode{LBtable} instructions}
\label{fig:lbtable}
\end{wrapfigure}

Unlike lookaheads, deciding whether a lookbehind is satisfied only depends on the part of the string that has already been read.
This suggests that we can merge together the first two steps of the algorithm in Section~\ref{sec:lookarounds}, building the oracle as we match the main expression.
As we synchronize both steps, it turns out that only the oracle column corresponding to the current string position is used.
This new algorithm leverages this observation.
It maintains an array of booleans, \texttt{LBtable} of length $\looknb{$r$}$.
This \texttt{LBtable} corresponds to the column of the oracle of Section~\ref{sec:lookarounds} of the current string position.
We show below how an NFA simulation engine can update this array in such a way that, at string position $i$, for a lookbehind $b$, \texttt{LBtable[$b$]} contains 1 if and only if the lookbehind $r\lkr{b}$ holds at position $i$.
When a thread encounters a lookbehind, one then simply needs to check the corresponding entry in \texttt{LBtable}.
We implement this with a new bytecode instruction, \bytecode{CheckLB b}, killing the current thread if \texttt{LBtable[b]} is 0.
For negative lookbehinds, we use a similar \bytecode{NegCheckLB b} instruction. 
Figure~\ref{fig:lbtable} shows how to augment Algorithm~\ref{alg:pikevm} for these new instructions.
At each new character, the \bytecode{LBtable} is also reset to an array of \bytecode{false}.
Compared to Figure~\ref{fig:writecheck}, the two instructions do not use the current string position \bytecode{i}, since the \bytecode{LBtable} corresponds only to the current column of the \bytecode{oracle}.

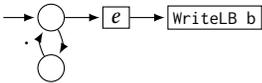
\begin{wrapfigure}{l}{3.8cm}
  \centering
\begin{tikzpicture}[%
        every node/.style={circle,minimum size=10pt,minimum height=2pt, inner sep=2pt},
        shorten >=1pt,
        node distance=0.5cm, >=latex,
        initial text={}
      ]
  \node [initial] (lbstart) [draw, xshift=-2cm] {};
  \node [] (loop) [draw, below=of lbstart,yshift=.2cm] {};
  \node [rectangle] (e) [draw, right=of lbstart] {\small$e$};
  \node [rectangle] (write) [draw, right=of e] {\scriptsize\bytecode{WriteLB b}};

  \path [draw] (lbstart) edge[->]  node {} (e);
  \path [draw] (e) edge[->]  node {} (write);
  \path [draw] (lbstart) edge[->, bend left] node {} (loop);
  \path [draw] (loop) edge[->,left,bend left] node {.} (lbstart);
\end{tikzpicture}
\caption{NFA of \rgx{(?<=\lkr{b}e)}}
\label{fig:lookbehind_nfa}  
\end{wrapfigure}

To update this \texttt{LBtable}, it suffices to compile each lookbehind separately and run them in lockstep with the main regex.
Each of these lookbehind automata ends with a new \bytecode{WriteLB b} bytecode instruction, writing to \texttt{LBtable} that the lookbehind $b$ is satisfied at the current string position.
Each lookbehind automaton also starts with a \rgx{.*?} prefix, allowing the match to begin at any character. 
Figure~\ref{fig:lookbehind_nfa} shows the NFAs generated for captureless lookbehinds; since thread priority is irrelevant in these automata, we do not use dashed arrows.
Each time the simulation reads a new character, the contents of \texttt{LBtable} are reset to 0.\footnote{One can even avoid resetting by storing the last position where each lookbehind was satisfied instead of booleans.}

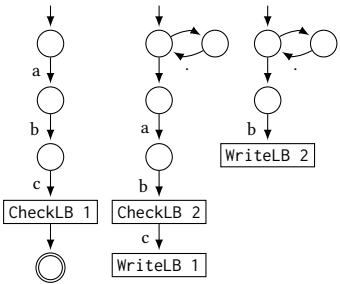
\begin{wrapfigure}{l}{5cm}
  \centering
  \scriptsize
\begin{tikzpicture}[%
        every node/.style={circle,minimum size=10pt,minimum height=2pt, inner sep=2pt},
        shorten >=1pt,
        node distance=0.38cm, >=latex,
        initial text={}
  ]
  \node [initial above] (maina) [draw] {};
  \node [] (mainb) [draw, below=of maina] {};
  \node [] (mainc) [draw, below=of mainb] {};
  \node [rectangle] (maincheck) [draw, below=of mainc] {\bytecode{CheckLB 1}};
  \node [accepting] (mainend) [draw, below=of maincheck] {};

  \node [initial above] (l1start) [draw, right=of maina, xshift=.7cm] {};
  \node [] (l1loop) [draw, right=of l1start] {};
  \node [] (l1a) [draw, below=of l1start] {};
  \node [] (l1b) [draw, below=of l1a] {};
  \node [rectangle] (l1check) [draw, below=of l1b] {\bytecode{CheckLB 2}};
  \node [rectangle] (l1write) [draw, below=of l1check] {\bytecode{WriteLB 1}};

  \node [initial above] (l2start) [draw, right=of l1start, xshift=.7cm] {};
  \node [] (l2loop) [draw, right=of l2start] {};
  \node [] (l2b) [draw, below=of l2start] {};
  \node [rectangle] (l2write) [draw, below=of l2b] {\bytecode{WriteLB 2}};

  \path [draw] (maina) edge[->, left]  node {a} (mainb);
  \path [draw] (mainb) edge[->, left]  node {b} (mainc);
  \path [draw] (mainc) edge[->, left]  node {c} (maincheck);
  \path [draw] (maincheck) edge[->]  node {} (mainend);
  \path [draw] (l1start) edge[->, bend left]  node {} (l1loop);
  \path [draw] (l1loop) edge[->, below, bend left]  node {.} (l1start);
  \path [draw] (l1start) edge[->]  node {} (l1a);
  \path [draw] (l1a) edge[->, left]  node {a} (l1b);
  \path [draw] (l1b) edge[->, left]  node {b} (l1check);
  \path [draw] (l1check) edge[->, left]  node {c} (l1write);
  \path [draw] (l2start) edge[->, bend left]  node {} (l2loop);
  \path [draw] (l2loop) edge[->, below, bend left]  node {.} (l2start);
  \path [draw] (l2start) edge[->]  node {} (l2b);
  \path [draw] (l2b) edge[->, left]  node {b} (l2write);
\end{tikzpicture}
\caption{Lookbehind automata for \regex{abc(?<=ab(?<=b)c)}}
\label{fig:memoryless_lookbehind}
\end{wrapfigure}

To run these multiple automata in lockstep, it suffices to add their initial state to the list of initial states of the NFA simulation.\footnote{Alternatively, since the lookbehinds considered in this algorithm do not contain groups, one could also run these lookbehind automata with a Lazy DFA matcher in lockstep with an NFA simulation.}
This list of initial states is ordered such that the initial state of a lookbehind comes before the initial state of a parent lookbehind in the regex AST, to ensure that each \bytecode{WriteLB} instruction is executed before the corresponding \bytecode{CheckLB} instruction.
As an example, consider the regex \regex{abc(?<=ab(?<=b)c)}.
Its NFA is shown on Figure~\ref{fig:memoryless_lookbehind}.
On the string \str{abc}, the NFA simulation will start executing all three automata, starting with the one on the right.
After reading \str{a}, none of the automata have yet reached a final state, and \texttt{LBtable} contains only 0s.
This is expected, since none of the lookbehinds hold after reading a single \str{a}.
After reading \str{b} however, the automaton on the right reaches the \bytecode{WriteLB 2} instruction, and writes to \texttt{LBtable}.
Just after this write, the middle automaton reaches the \bytecode{CheckLB 2} instruction, and since \texttt{LBtable[2]} was just written to, this thread is kept alive.
Similarly after reading \str{c}, the middle automaton will write to the table and the main automaton on the left will finally reach an accepting state, indicating that a match was found.

\texttt{LBtable} has size $\looknb{r}$, bounded by $\bigo{\size{r}}$.
Since each lookbehind is compiled exactly once, the total generated bytecode has size $\bigo{\size{r}}$.
Each \bytecode{WriteLB} and \bytecode{CheckLB} instruction can be executed with $\bigo{1}$ complexity,
and resetting \texttt{LBtable} at each new character has a total time complexity of $\bigo{\size{r}\times\size{s}}$, leaving the total complexity of the NFA simulation unchanged.

\paragraph{Correctness}
This algorithm follows from Observation 1 and the following property:
for any lookbehind \regex{(?<=r)}, any string $s$ and index $i$, there exists a match of \regex{.*?r} on $s$ ending at position $i$ if and only if \regex{(?<=r)} holds in string $s$ at position $i$.

\subsection{Linear Matching of the JavaScript Plus}
\label{sec:linear_plus}

\begin{wrapfigure}{l}{3.5cm}
\small
\begin{tikzpicture}[%
        every node/.style={circle,minimum size=10pt,minimum height=2pt, inner sep=2pt},
        shorten >=1pt,
        node distance=0.5cm, >=latex,
        initial text={}
      ]
  \node [initial] (p1) [draw] {};
  \node [rectangle] (p2) [draw, right=of p1] {$e$};
  \node [] (p3) [draw, right=of p2] {};
  \node [] (p4) [right=of p3] {};
  \node [initial] (lp1) [draw, below=of p1,yshift=0cm] {};
  \node [rectangle] (lp2) [draw, right=of lp1] {$e$};
  \node [] (lp3) [draw, right=of lp2] {};
  \node [] (lp4) [right=of lp3] {};
  \path [draw] (p1) edge[->]  node {} (p2);
  \path [draw] (p2) edge[->]  node {} (p3);
  \path [draw] (p3) edge[->]  node {} (p4);
  \path [draw] (p3) edge[->,dotted,bend left=40]  node {} (p1);
  \path [draw] (lp1) edge[->]  node {} (lp2);
  \path [draw] (lp2) edge[->]  node {} (lp3);
  \path [draw] (lp3) edge[->,dotted]  node {} (lp4);
  \path [draw] (lp3) edge[->,bend left=40]  node {} (lp1);
\end{tikzpicture}
\caption{Usual NFA for \rgx{e+} and \rgx{e+?} in linear engines.}
\label{fig:usual_plus}
\end{wrapfigure}
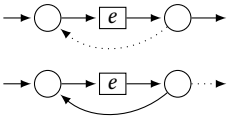

The traditional construction for a regex plus (as shown in Figure~\ref{fig:usual_plus} and used in RE2 or Rust) generates linear-sized NFAs.
Unfortunately, this construction does not implement the JavaScript plus semantics.
To see where this construction fails, consider the regex \regex{\noncap{\eps|.}+} on string \str{a}.
In JavaScript, an engine returns a match on the whole string, doing a first iteration matching nothing then another matching \str{a}.
With the usual construction of Figure~\ref{fig:usual_plus} however, an NFA simulation's highest priority match does a single iteration matching only the empty string.
Doing another iteration is invalid, since it would go back to the beginning of the plus without having consumed any character.
While this is reminiscent of the problem solved in Section~\ref{sec:nullable}, wrapping the body in \bytecode{BeginLoop} and \bytecode{EndLoop} instructions would not be correct, as it would prevent a plus from ever matching the empty string.
As seen in Section~\ref{subsec:js_semantics}, the first mandatory repetition of the plus can match the empty string, but not the following ones.
To reflect this difference, the solution implemented by V8Linear was simply to expand \regex{e+} into \regex{ee*} and \regex{e+?} into \regex{ee*?}.
However, this leads to regex-exponential complexity:
for instance, \regex{\noncap{\noncap{a+}+}+} gets expanded into \regex{\noncap{\noncap{aa*}+}+}, then \regex{\noncap{\noncap{aa*}\noncap{aa*}*}+} and \regex{\noncap{\noncap{aa*}\noncap{aa*}*}\noncap{\noncap{aa*}\noncap{aa*}*}*}.
To prevent exponential explosion, V8Linear used to simply reject regexes with too many nested plusses.
In this section, we present two new constructions to implement any nonnullable or greedy plus without such bytecode duplication and in linear time.

\def\nullable{\ensuremath{\operatorname{null}}}
\def\nullor{\ensuremath{\operatorname{max}}}
\def\nulland{\ensuremath{\operatorname{min}}}
\def\nn{\textbf{NN}}
\def\cin{\textbf{CIN}}
\def\cdn{\textbf{CDN}}
\begin{wrapfigure}{l}{4.6cm}
  \small
\begin{tabular}{c}
  $\nullable(.)=\nullable(c)$ = \nn\\
  $\nullable(\eps)$ = \cin\\
  $\nullable(e_1~e_2)$ = $\nullor(\nullable(e_1),\nullable(e_2))$\\
  $\nullable(e_1|e_2)$ = $\nulland(\nullable(e_1),\nullable(e_2))$\\
  $\nullable(e \rgx{*}) = \nullable(e \rgx{*?})$ = \cin\\
  $\nullable(e \rgx{+}) = \nullable(e \rgx{+?})$ = $\nullable(e)$\\
  $\nullable((e))$ = $\nullable(e)$\\
  $\nullable(\mathit{lk}~e)$ = \cdn\\
\end{tabular}

Using the order \nn~$>$~\cdn~$>$~\cin.\\
\caption{Nullability analysis}
\label{fig:nullability}
\end{wrapfigure}

\subsubsection{Nullability}
We now define three notions of nullability.
A regex can be nonnullable (or NN), meaning that it cannot ever match the empty string (\emph{e.g.} \regex{a}).
Otherwise, we define \textit{Context-Independent Nullable} (CIN) and \textit{Context-Dependent Nullable} (CDN) regexes.
A CIN can always match the empty string (\emph{e.g.} \regex{a|\eps}).
A CDN however can match the empty string depending on the surrounding context
(when there are lookarounds or if the engine supports anchors like \rgx{\textbackslash b} or \rgx{\^}).
For instance, \regex{a|(?=b)} is only nullable when the next character is a \str{b}.
Figure~\ref{fig:nullability} presents a syntax-directed nullability analysis.
This is an approximation, in the sense that a nonnullable or a CIN can be classified as CDN.
As the analysis does not explore lookarounds it does not detect for instance, that \regex{a|(?=b)(?!b)} is nonnullable or that \regex{(?=a)|(?!a)} is a CIN.
This isn't an issue: the solution we present for CDNs is also correct for CINs and nonnullables, albeit more complex.

\subsubsection{NonNullable Plus}
The simplest case happens when compiling \regex{e+} or \regex{e+?} where \rgx{e} is nonnullable.
In that case, it turns out that the usual NFA construction shown in Figure~\ref{fig:usual_plus} correctly implements the JavaScript semantics.
Since \rgx{e} cannot match the empty string, it does not matter that the first repetition is allowed to match it while the others are not.
We have implemented and merged this case in V8Linear. 
For counted repetitions \regex{e\{n,\}}, when $n>0$ and \rgx{e} is nonnullable we can similarly avoid one repetition of the bytecode of \rgx{e}
by replicating it $n-1$ times, followed by Figure~\ref{fig:usual_plus}, instead of the usual $n$ repetitions followed by \rgx{e*}.

\subsubsection{CIN and CDN Plus}
\label{subsec:cin_cdn}

We now present a way to match nullable greedy plusses linearly.
Matching nullable \textit{lazy} plusses in linear time remains an open problem, although we only found them in 0.003\% of regexes (see Section~\ref{subsec:stats}).
Our solution leverages the following observation:
the only way to match the empty string with a greedy nullable JavaScript plus is to do a single iteration, and this has the \emph{lowest priority}.
For instance, consider the regex \regex{\noncap{a|\eps|b}+}, a CIN.
First, the middle branch of the plus body is only allowed in the first iteration of the plus.
So we can expand and rewrite this regex as \regex{\noncap{a|\eps|b}\noncap{a|b}*}, removing the empty path from the star.
This regex can only match the empty string if it skips the star, but doing so has the lowest priority since the star is greedy.
Even if the string starts with a \str{b}, then matching empty in the first iteration then consuming \str{b} in the greedy star has higher priority than just consuming empty in the first iteration and skipping the star.
We can then further rewrite the regex to \regex{\noncap{a|b}\noncap{a|b}*|\eps}, or even \regex{\noncap{a|b}+|\eps}, where the plus is now nonnullable.

\begin{wrapfigure}{l}{3.3cm}
\scriptsize
\begin{tikzpicture}[%
        every node/.style={circle,minimum size=10pt,minimum height=2pt, inner sep=2pt},
        shorten >=1pt,
        node distance=0.25cm, >=latex,
        initial text={}
      ]
  \node [initial] (entry) [draw] {};
  \node [] (nonnull) [draw, below left=of entry, yshift=-.2cm] {};
  \node [rectangle] (setquant) [draw, below=of nonnull] {\bytecode{SetQuant}};
  \node [rectangle] (begin) [draw, below=of setquant] {\bytecode{BeginLoop}};
  \node [rectangle] (body) [draw, below=of begin] {$e$};
  \node [rectangle] (end) [draw, below=of body] {\bytecode{EndLoop}};
  \node [] (fork) [draw, below=of end] {};
  \node [rectangle] (quant) [draw, right=of body, xshift=.2cm] {\bytecode{SetNullPlus}};
  \node [rectangle] (check) [draw, above=of quant] {\bytecode{CheckNull}};
  \node [accepting] (exit) [draw, below right=of fork, yshift=-.2cm] {};

  \path [draw] (entry) edge[->, dotted]  node {} (nonnull);
  \path [draw] (nonnull) edge[->]  node {} (setquant);
  \path [draw] (setquant) edge[->]  node {} (begin);
  \path [draw] (begin) edge[->]  node {} (body);
  \path [draw] (body) edge[->]  node {} (end);
  \path [draw] (end) edge[->]  node {} (fork);
  \path [draw] (fork) edge[->, dotted]  node {} (exit);
  \path [draw] (entry) edge[->,bend left=20]  node {} (check);
  \path [draw] (check) edge[->]  node {} (quant);
  \path [draw] (quant) edge[->,bend left=20]  node {} (exit);
  \path [draw] (fork) edge[->,bend left=60]  node {} (nonnull);
\end{tikzpicture}
\caption{NFA for a CDN \rgx{e+}}
\label{fig:nfa_cdn_cin}
\end{wrapfigure}
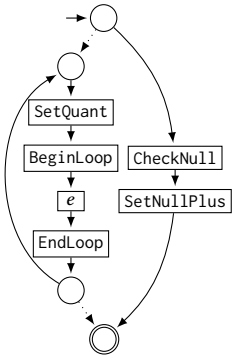

As a result, we compile CINs and CDNs as a \bytecode{Fork}, where the left branch contains the nonnullable paths, and the right one corresponds to matching the empty string.
However, the rewriting transformation in the example above cannot be generalized.
First, extracting the nullable path from a regex cannot always be done this easily (consider \regex{\noncap{\noncap{a|\eps}\noncap{b|\eps}}+}).
Second, for CDNs, the nullable path should only be taken when the current string position allows it.
Finally, the nullable path of a regex may define capture groups.
These groups can either be empty groups (\emph{e.g.} in \regex{\noncap{a|(\eps)|b}+}), or nonempty groups if the regex has lookarounds (\emph{e.g.} the CIN \regex{\noncap{a|(?=(c))|\eps|b}+}).

The first issue can be dealt with by adding \bytecode{BeginLoop} and \bytecode{EndLoop} instructions (see Section~\ref{sec:nullable}) around the body of the plus, effectively removing the nullable paths.
The second issue for CDNs can be solved by adding a new \bytecode{CheckNull} instruction on the nullable path checking that the nulled plus is in fact nullable at that point.
We leave implementation details out of this explanation, but computing the nullability of all plusses (even nested) in a regex $r$ for a particular string position can be done with time complexity $\bigo{\size{r}}$, for a total added complexity of $\bigo{\size{r}\times\size{s}}$.\footnote{For instance, by memoizing checks for the nullability of nested CDN plusses.}

Finally, when taking the nulled path of \regex{e+}, one needs to set the capture groups defined along the top-priority nullable path in \rgx{e} at that string position.
This problem is similar to reconstructing capture groups inside lookarounds in Section~\ref{sec:lookarounds}.
From the Capture Reset property, we know that these capture groups are only defined if the last time the plus was matched, the winning thread went through its nullable path.
Like for lookarounds, we record the last string position at which each plus was nulled with a new bytecode instruction \bytecode{SetNullPlus} inserted along the nullable path.
The resulting linear-sized NFA for \regex{e+} can be seen on Figure~\ref{fig:nfa_cdn_cin}.
Like a \bytecode{SetQuant} instruction, this instruction records in the current thread memory the current clock (see Section~\ref{sec:clocks}), but it also records that the last time this plus was matched, it was nulled and at which string position.
After a winning thread is found, one can reconstruct the missing groups a posteriori.
To do so, one starts the simulation again on the body of each nulled plus, at the string position where they were last nulled, without consuming any character of the string.
One only needs to run this other simulation once per CIN or CDN of the main expression \rgx{r} just like for the lookarounds of the third stage of Section~\ref{sec:lookarounds} but only the empty string.
This last phase has a total time complexity of $\bigo{\size{\rgx{r}}}$,

\begin{wrapfigure}{L}{3.0cm}
  \begin{minipage}{3.0cm}
\begin{algorithm}[H]\scriptsize
\SetKwFor{case}{case}{$\Rightarrow$}{}
\SetKw{hlet}{let}
\SetInd{0.1em}{0.75em}
\SetAlgoLined
\case{\bytecode{SetQuant}}
     {\texttt{t.qclocks[r] = clk}\\
       \texttt{t.plusnulled[q] = -1}\\
       \texttt{t.pc = t.pc+1}}
\case{\bytecode{SetNullPlus q}}
     {\texttt{t.qclocks[q]=clk}\\
       \texttt{t.plusnulled[q] = i}\\
       \texttt{t.pc = t.pc + 1}}
\case{\bytecode{CheckNull q}}
     {\eIf{\texttt{nullable(q,i)}}
       {t.pc = t.pc + 1}
       {active.pop()}}
\end{algorithm}
  \end{minipage}
\caption{Nullable plus instructions}
\label{fig:nullinstr}
\end{wrapfigure}

Figure~\ref{fig:nullinstr} shows how to extend Algorithm~\ref{alg:pikevm} for these new instructions.
Threads contain an additional set of registers, \bytecode{plusnulled}, indicating for each quantifier \bytecode{q} if its last iteration consisted in taking the nullable path of a nullable plus (in which case it contains the string position when it last happened), or not (in which case it contains \bytecode{-1}).
These registers are used at the end of the match to figure out which CDN plus needs its capture groups to be reconstructed.
We modify the \bytecode{SetQuant} instruction of Figure~\ref{fig:clk} so that this register is reset to \bytecode{-1} when the quantifier is taken without going through the nullable path.
The \bytecode{nullable(q,i)} function checks the nullability of the \bytecode{q} quantifier at the current string position \bytecode{i}. When \bytecode{q} contains lookarounds, this reads from the oracle of Section~\ref{sec:lookarounds}.

This algorithm also supports nested CIN or CDN plusses with capture groups inside.
Consider for instance the regex \regex{((\eps)+$_2$)+$_1$} being matched on the string \str{a}.
After doing the NFA simulation, a match is found where the winning thread took the nullable path of the outer plus, \rgx{+$_1$}.
In this thread, \bytecode{plusnulled[1]} is \bytecode{0}, so we reconstruct the capture groups inside \rgx{+$_1$} at string position \bytecode{0}.
To do so, we interpret \rgx{((\eps)+$_2$)}
\footnote{For this step, as we reconstruct groups that were taken along a nullable path, we can omit the non-nullable path on the left of Figure~\ref{fig:nfa_cdn_cin} for nested CIN/CDN plusses. This ensures linearity: each plus is compiled at most once during this phase.}.
Once again, a match is found where the outer capture group has been defined, and now in the winning thread \bytecode{plusnulled[2]} is \bytecode{0}.
Consequently, we then interpret \rgx{(\eps)} from string position \bytecode{0}, defining the inner capture group.

\paragraph{Correctness}
The new NFA construction relies on the fact that matching the empty string with a greedy plus has the lowest priority:
if at the current position, the regex inside the plus can match any other string $s$ with lower priority, then matching empty in the first iteration then matching $s$ in a second iteration has more priority than doing a single empty iteration since the plus is greedy.
The capture group reconstruction algorithm uses the Capture Reset property: any capture group in a plus can only be defined by the last iteration of that plus in the winning thread. The \bytecode{SetNullPlus} instruction ensures that missing groups are reconstructed.

\subsection{A Space-Time Complexity Tradeoff for Capture Groups in an NFA Simulation}
\label{sec:tradeoff}

\begin{wrapfigure}{l}{6cm}
  \small
\begin{tabular}{r@{\,}|@{\,}c@{}c}
  & Time & Space\\
  \hline
  Array & $\bigo{\size{r}^2\times\size{s}}$ & $\bigo{\size{r}^2}$ \\
  Linked List & $\bigo{\size{r}\times\size{s}}$ & $\bigo{\size{r}\times\size{s}}$ \\
  Balanced Tree & $\bigo{\size{r}\times\mathit{log}\size{r}\times\size{s}}$ & $\bigo{\size{r}^2}$\\
\end{tabular}
\caption{NFA simulation complexity using different thread register data-structures.}
\label{fig:tradeoff}
\end{wrapfigure}

Recall the execution of the \bytecode{Fork} instruction in Algorithm~\ref{alg:pikevm}.
As threads contain string indices for each capture group (and the clock values of Section~\ref{sec:clocks}), this instruction requires copying $\bigo{\size{r}}$ data in the worst case.
All the NFA simulation implementations we found (including Rust, RE2 and V8Linear) use an array for thread registers and copy the array when executing a \bytecode{Fork}.
As a result, the entire simulation execution can have worst-case time complexity $\bigo{\size{r}^2\times\size{s}}$.
In most practical cases where the string is significantly bigger than the regex, this is not an issue.
However, in the case of user-provided regexes, this quadraticity can become a security concern.
We present two alternative data-structures for thread registers.

If one needs strict regex-size linearity, one can store each update done to a thread's registers in an immutable linked list.
Storing a new value (\bytecode{SetReg}) then consists in adding a new cell to the list.
Threads can be forked in $\bigo{1}$ time by simply sharing the tail of the list across multiple threads.
Threads registers are often written to during execution, but only read from at the very end of the match.
Extracting the final value of each register for the winning thread can be done by traversing its list of updates.
This comes with an additional space complexity for bigger strings: we know that there can be as many as $\bigo{\size{r}\times\size{s}}$ executions of \bytecode{SetReg} instructions in a match and as a result as many allocations.
\footnote{This can be seen as a specialization of the persistent array implementation of~\cite{persistent_ds}, where the \texttt{Diff} list is the list of register updates and a single \textit{rerooting} is performed at the end for the winning thread.}

If an $\bigo{\size{r}\times\mathit{log}\size{r}\times\size{s}}$ time complexity is acceptable, one can instead store thread registers in an immutable balanced tree.
Forking is then achieved in $\bigo{1}$ time.
However, executing \bytecode{SetReg} instructions now requires $\bigo{\mathit{log}(\size{r})}$ time.
The worst-case space complexity of $\bigo{\size{r}^2}$ is achieved when $\bigo{\size{r}}$ threads are alive without sharing any sub-trees.

\begin{wrapfigure}{l}{.25\textwidth}
  \includegraphics[width=.25\textwidth]{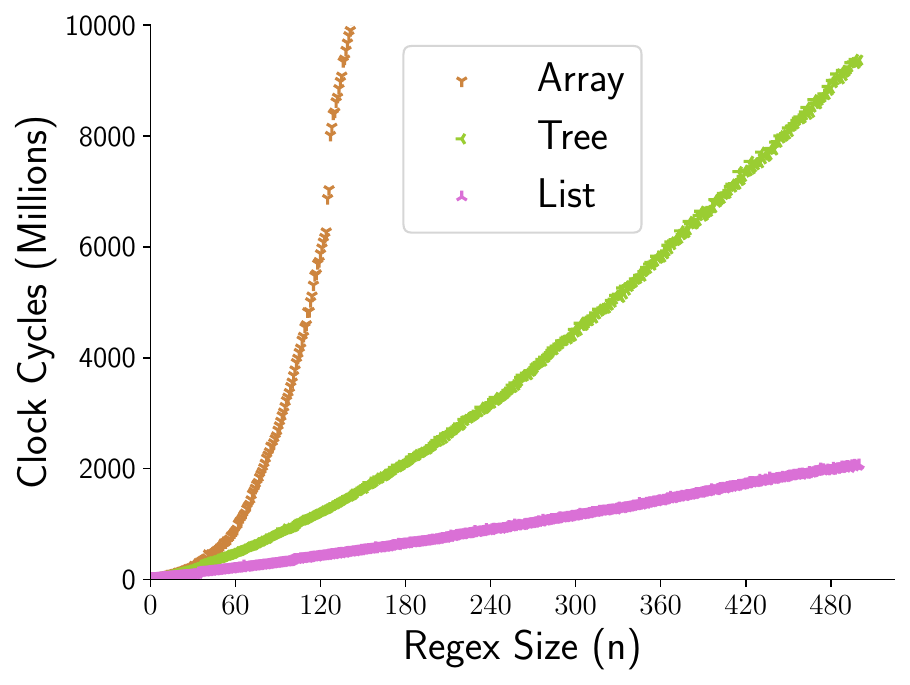}%
  \caption{$r_n$ matching time}
  \label{fig:bench_tradeoff}
\end{wrapfigure}
  
We experimentally validated that RE2 and Rust exhibit such regex-quadratic complexity, although Rust limits the maximum number of capture groups in a regex.
We implemented these three solutions in our prototype OCaml engine and experimentally validated their time complexities.
For instance, we define the following regex family: $r_0=$\regex{\noncap{(a)?}*}, $r_1=$\regex{\noncap{(a)?(a)?}*}, $r_2=$\regex{\noncap{(a)?(a)?(a)?}*} and so on. 
Figure~\ref{fig:bench_tradeoff} shows the execution time, measured in CPU clock cycles, of matching $r_n$ on the string \str{a$^{1000}$}.
The experimental setup is described in Section~\ref{subsec:exps}.
We believe that regex engines would benefit from simple heuristics, for instance switching to a linked list implementation if the number of capture groups is bigger than the size of the input string.

\subsection{Composing Our NFA Simulation Extensions}
\label{sec:composition}

All the JavaScript regex features showcased on Figure~\ref{fig:syntax} can be supported together with linear time and space guarantees.
The algorithms we presented in Sections~\ref{sec:nullable}, \ref{sec:clocks}, \ref{sec:lookarounds}, \ref{sec:linear_plus} are designed to be composed together.
The NFA simulation algorithm~\ref{alg:pikevm} can be extended with Figures~\ref{fig:beginend}, \ref{fig:clk}, \ref{fig:writecheck} and \ref{fig:nullinstr}.
Three of these algorithms perform matching in several passes: both the algorithms of Sections~\ref{sec:lookarounds} and \ref{sec:linear_plus} reconstruct capture groups a posteriori, and the algorithm of Section~\ref{sec:clocks} filters capture groups after a match is found.
These different phases can be organized as follows.
In the first phase of Section~\ref{sec:lookarounds}, when building the oracle, there is no need to reconstruct the groups inside plusses since the oracle does not remember any capture group information.
Next, in the second phase, we first match the main expression, then reconstruct the groups inside nulled CDN plusses as in the second phase of Section~\ref{sec:linear_plus}.
Finally in the third phase of Section~\ref{sec:lookarounds}, we can match each lookaround subexpression, then reconstruct the capture groups inside each nulled CDN plus.
The clock filtering can be done at the very end, once a final match with all captures is found.

Each of the three register implementations of Section~\ref{sec:tradeoff} are compatible with all the algorithms presented in this paper.
While the streaming algorithm of Section~\ref{sec:memoryless} can be composed with the algorithms of Section~\ref{sec:nullable} and \ref{sec:clocks}, it cannot be composed with the algorithm of Section~\ref{sec:linear_plus} for CDNs if the CDNs contain capture groups.
In our CDN algorithm, we need to evaluate the nullability of each CDN plus, and this may depend on knowing if a lookaround holds at a previous string position, for which the full oracle is needed.

\section{Evaluation}
\label{sec:eval}

\subsection{Regex Usage Statistics}
\label{subsec:stats}

\begin{wrapfigure}{L}{6cm}
  \small
\begin{tabular}{l | c c}
  Nullable Quantifiers & 5729 & 0.37\% \\
  Capture in Quantifiers & 102283 & 6.67\% \\
  Nonnullable Plus (\rgx{+} or \rgx{+?}) & 405179 & 26.38\% \\
  CIN and CDN greedy \rgx{+} & 1041 & 0.07\% \\
  CIN and CDN lazy \rgx{+?} & 50 & 0.003\% \\
  Lookarounds & 79754 & 5.19\% \\
  Captureless Lookbehinds & 22734 & 1.48\% \\
\end{tabular}
\caption{Number of regexes using each feature}
\label{fig:stats}
\end{wrapfigure}

How often do developers use each of the features for which we presented new algorithms?
To answer this question, we parsed and analyzed large corpora of regexes from previous related work~\cite{lingua_franca,redos_impact}.
These consists of regexes scrapped online on StackOverflow, RegexLib, NPM or Pypi packages and others.
Some of these regexes may be written for other regex languages (like Python), but the syntax is mostly similar for the features that we studied.
We parsed them all with a JavaScript regex parser we wrote in OCaml, following the ECMA grammar~\cite{ecma_262} but rejecting unsupported features like backreferences.
In total, we parsed and analyzed 1536196 out of 1755587 regexes (87.5\%).
Figure~\ref{fig:stats} reports how many regexes include each feature.

Note that not all nullable quantifiers can exhibit the semantic mismatch presented in Section~\ref{sec:nullable}.
For instance, we believe that \regex{\noncap{a?b?}*} would return the same result in both semantics for any input string.
Finding a more precise syntactic characterization of quantifiers that require \bytecode{BeginLoop/EndLoop} instructions is left as future work.
For regexes with capture groups inside quantifiers, our algorithm of Section~\ref{sec:clocks} reduces the bytecode size.
More than a fourth of regexes include a nonnullable plus, for which previous techniques unnecessarily duplicate bytecode.
The more elaborate CIN and CDN technique of Section~\ref{subsec:cin_cdn} is only required for a thousand regexes, and only 50 use a nullable lazy plus, for which we haven't found a linear algorithm yet.
Around 5\% of all regexes use lookarounds of any kind, and more than a fourth of them only use captureless lookbehinds and could be handled by our algorithm without any additional memory complexity.

\subsection{Implementation}
\label{subsec:implem}

Each of our algorithms has been implemented in a standalone OCaml NFA simulation engine of around 3.5K lines of code (of which 350 are for parsing), available as an artifact.
This engine supports all the features presented here, as well as character classes, backslash sequences, anchors and counted repetition.
However regex flags, named groups, Unicode, hexadecimal and octal escapes have not yet been implemented, although they do not represent additional algorithmic complexity.
Backreferences are not supported.
The three different capture register implementations of Section~\ref{sec:tradeoff} can be used.
The engine comes with a differential fuzzer that randomly generates regexes and strings and compares the result to Irregexp.
This helped us discover the semantic bug of Section~\ref{sec:nullable}.
Since then, all our algorithms went through hours of fuzzing and millions of tests without reporting any mismatch.

We implemented several of our algorithms in V8Linear.
Three of them have been merged into V8Linear:
the semantic adaptation for nullable quantifiers (Section~\ref{sec:nullable}), the linear construction for nonnullable plus (Section~\ref{sec:linear_plus}), and the captureless lookbehind algorithm (Section~\ref{sec:memoryless}).
These three algorithms have been reviewed by V8 maintainers and tested against various JavaScript test suites, including browser test suites and the official Test262 JavaScript conformance test suite~\cite{test_262}.
We also contributed some of our own tests to the V8 test suite covering our contributions and documenting previous bugs of V8Linear.
We are currently working on implementing the other algorithms, starting with the clock algorithm for Capture Reset (Section~\ref{sec:clocks}).


\subsection{Experimental validation of complexity claims}
\label{subsec:exps}

\def\graphwidth{.25\textwidth}
\def\minigraphwidth{.5\textwidth}
\begin{figure}
  \begin{minipage}{.49\textwidth}
  \includegraphics[width=\minigraphwidth]{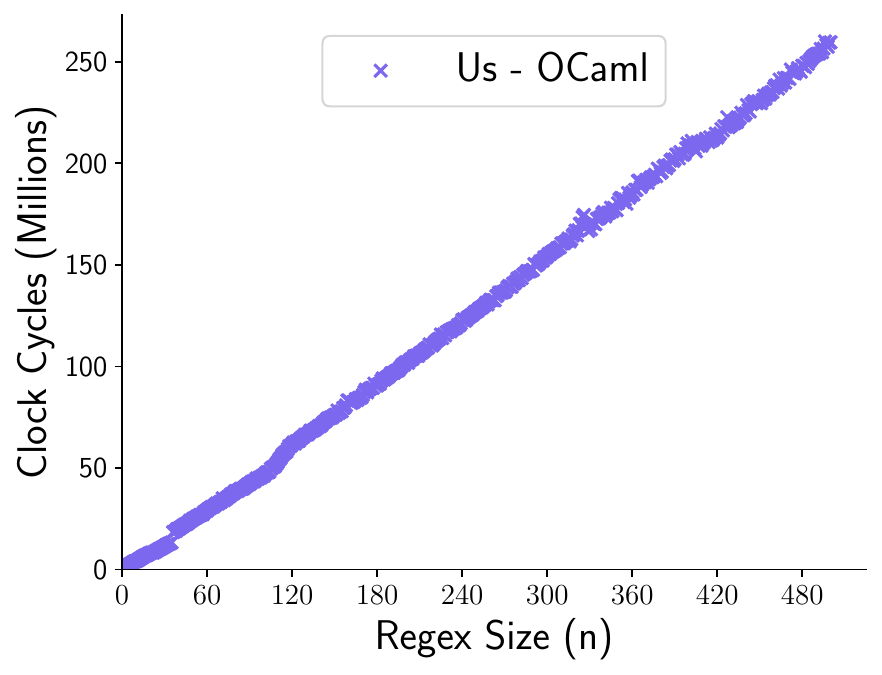}%
  \includegraphics[width=\minigraphwidth]{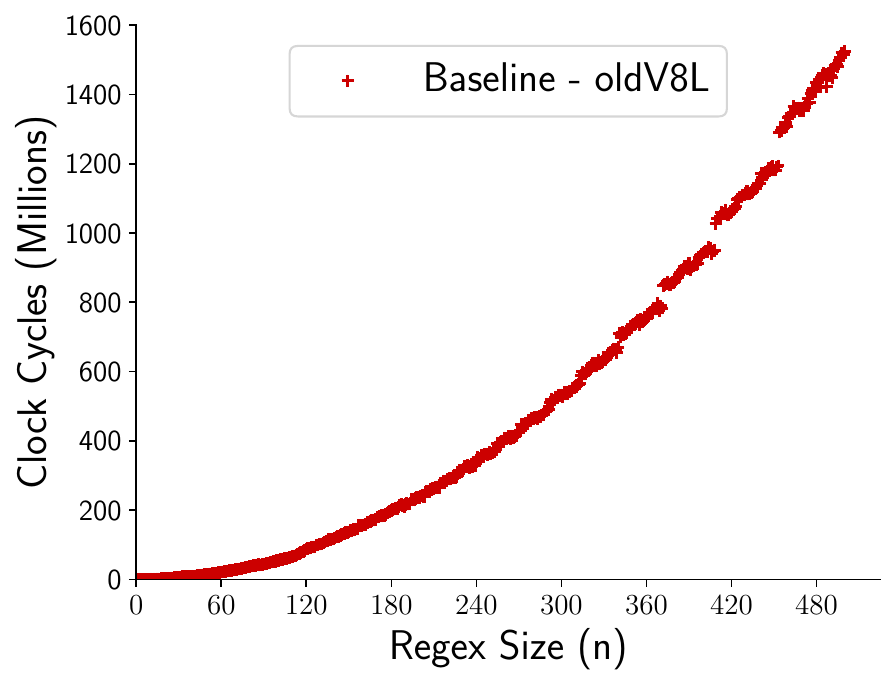}%
  \vspace{-0.3cm}%
  \caption{Capture Reset Complexity}
  \label{fig:bench_capture}
  \end{minipage}%
  \hfill\vline\hfill
  \begin{minipage}{.49\textwidth}
  \includegraphics[width=\minigraphwidth]{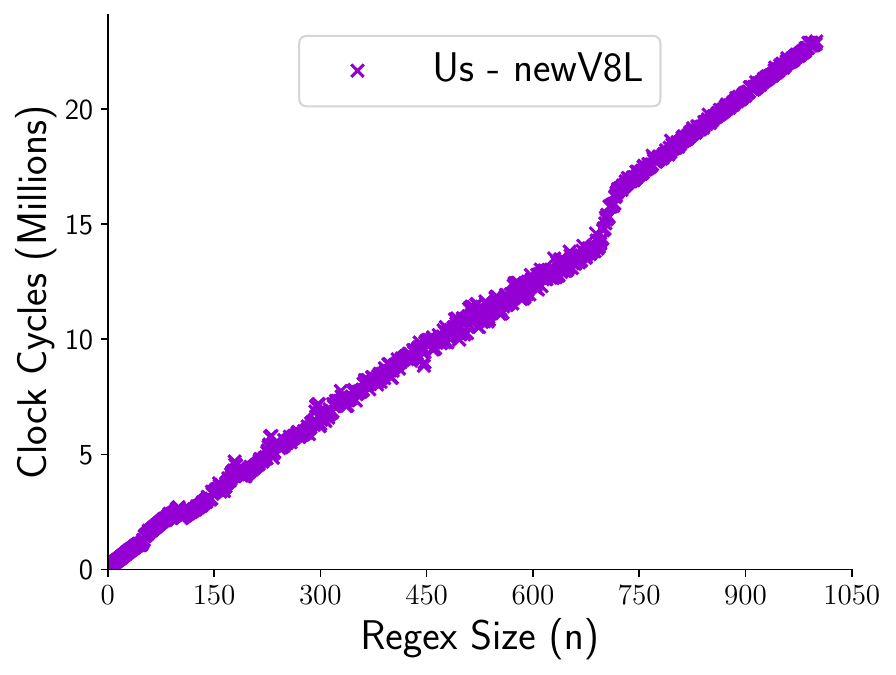}%
  \includegraphics[width=\minigraphwidth]{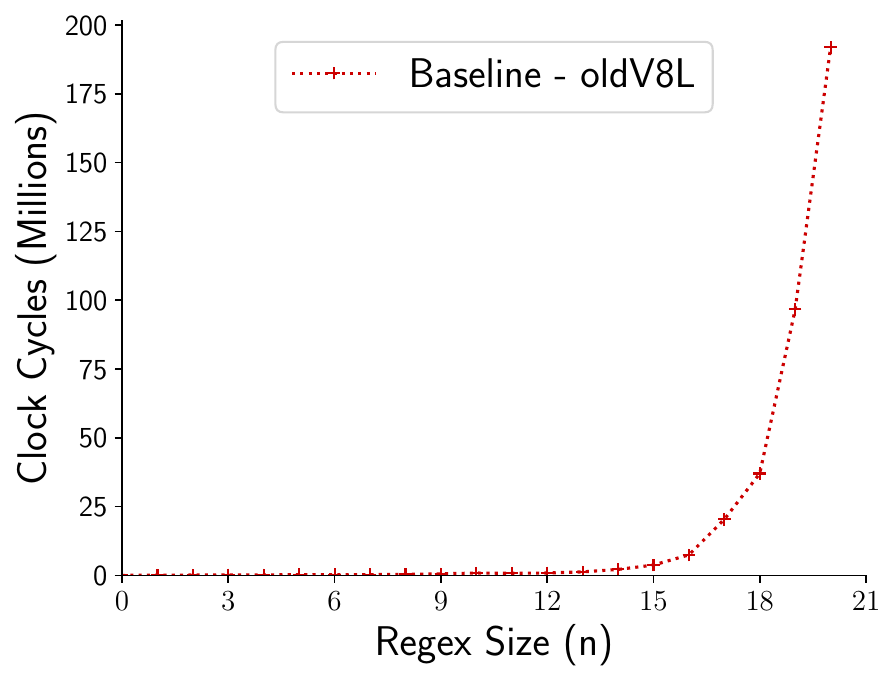}%
  \vspace{-0.3cm}%
  \caption{NonNullable Plus Complexity}
  \label{fig:bench_nn}
  \end{minipage}
  
  \begin{minipage}{.49\textwidth}
  \includegraphics[width=\minigraphwidth]{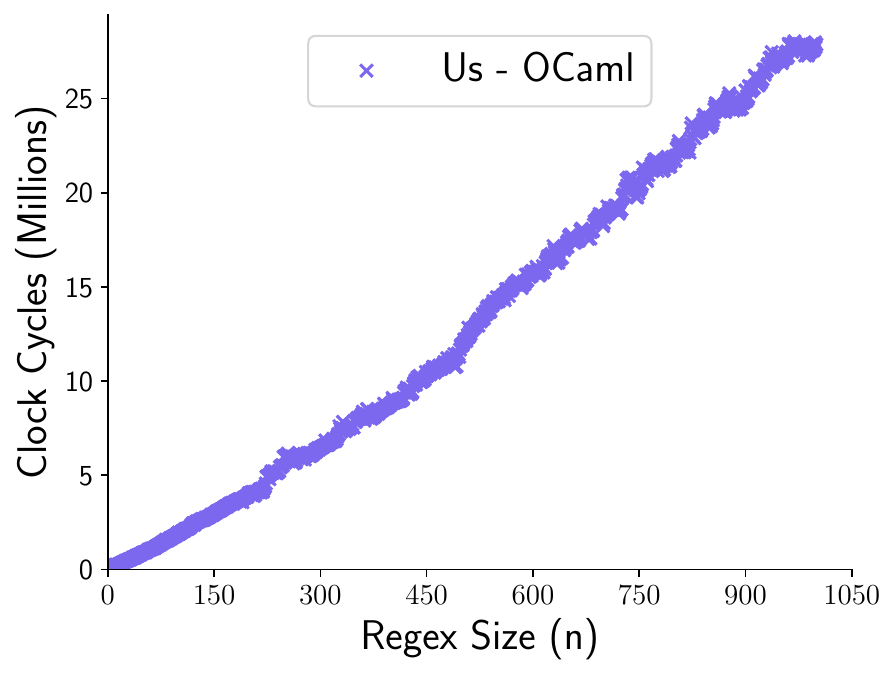}%
  \includegraphics[width=\minigraphwidth]{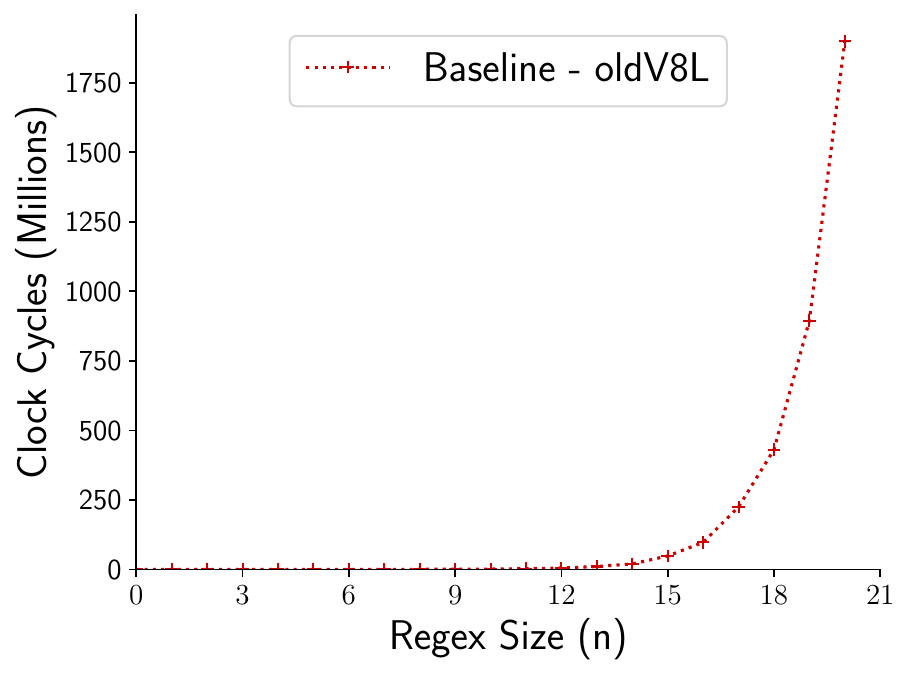}%
  \vspace{-0.3cm}%
  \caption{CDN Complexity}
  \label{fig:bench_cdn}
  \end{minipage}%
  \hfill\vline\hfill
  \begin{minipage}{.49\textwidth}
  \includegraphics[width=\minigraphwidth]{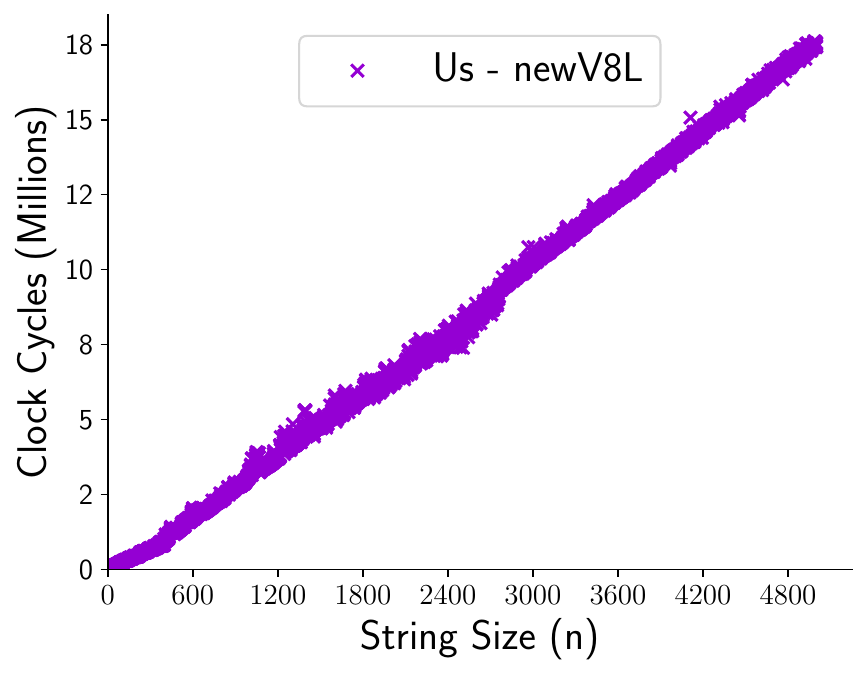}%
  \includegraphics[width=\minigraphwidth]{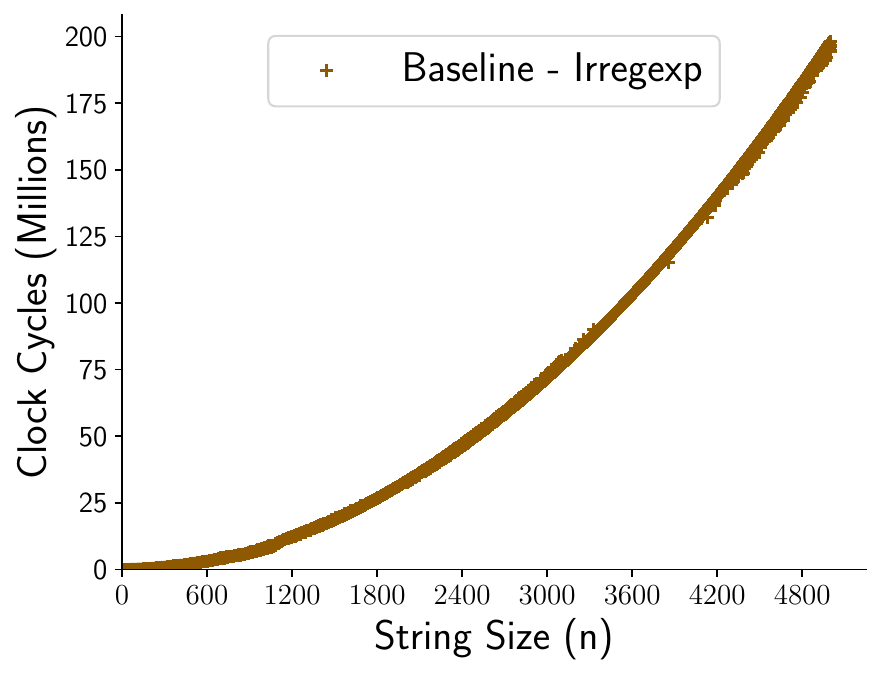}%
  \vspace{-0.3cm}%
  \caption{Captureless Lookbehind Complexity}
  \label{fig:bench_lookbehind}
  \end{minipage}

  \begin{minipage}{.49\textwidth}
  \includegraphics[width=\minigraphwidth]{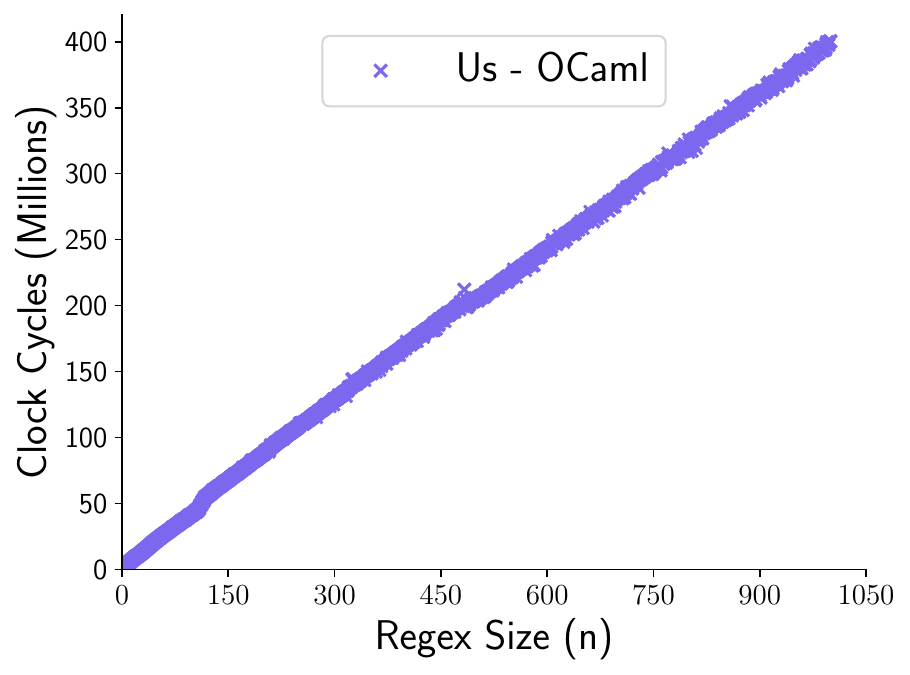}%
  \includegraphics[width=\minigraphwidth]{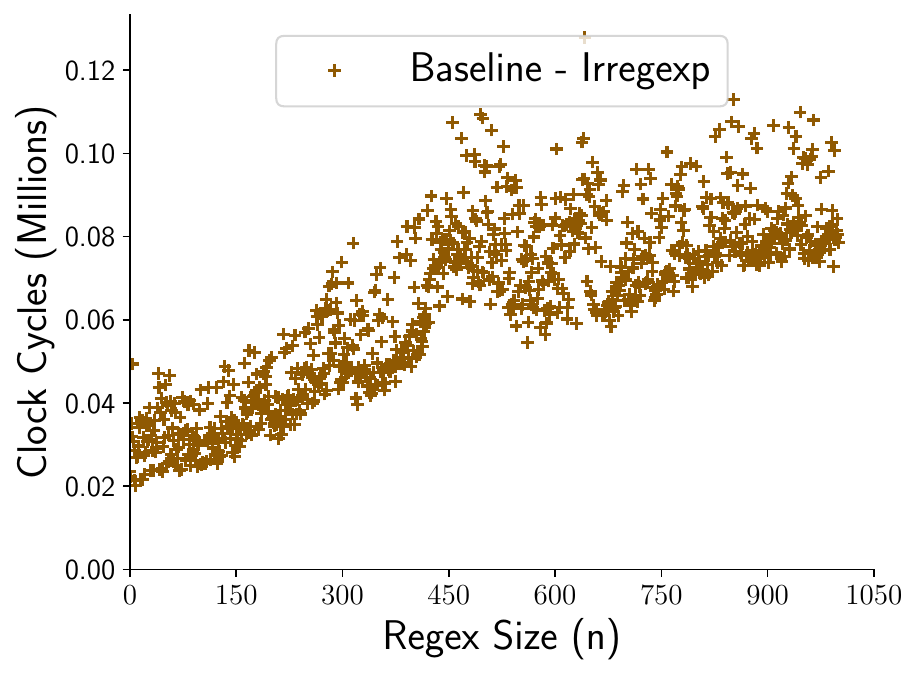}%
  \vspace{-0.3cm}%
  \caption{Lookarounds Regex-Complexity}
  \label{fig:bench_lareg}  
  \end{minipage}%
  \hfill\vline\hfill
  \begin{minipage}{.49\textwidth}
  \includegraphics[width=\minigraphwidth]{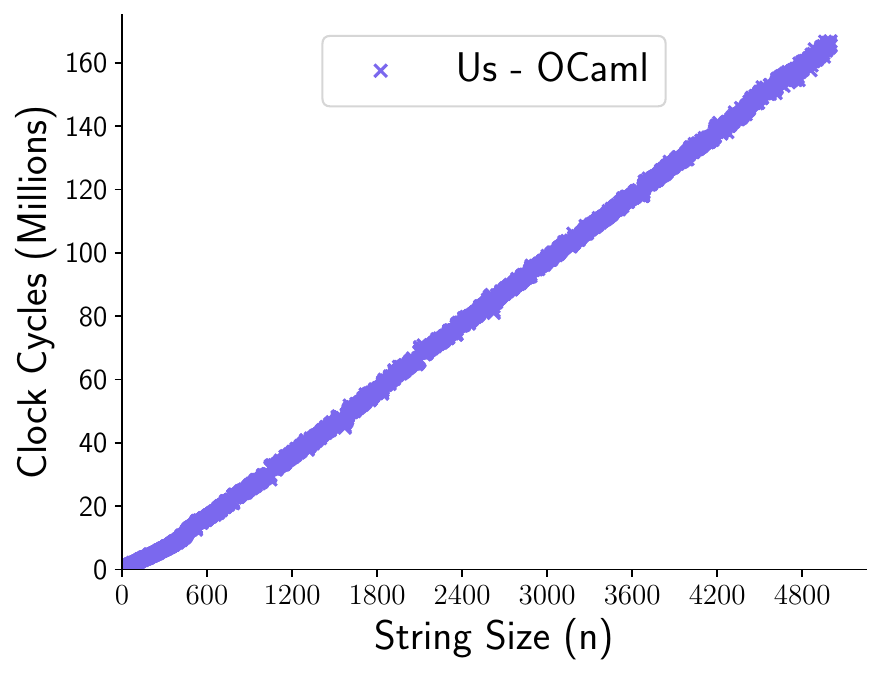}%
  \includegraphics[width=\minigraphwidth]{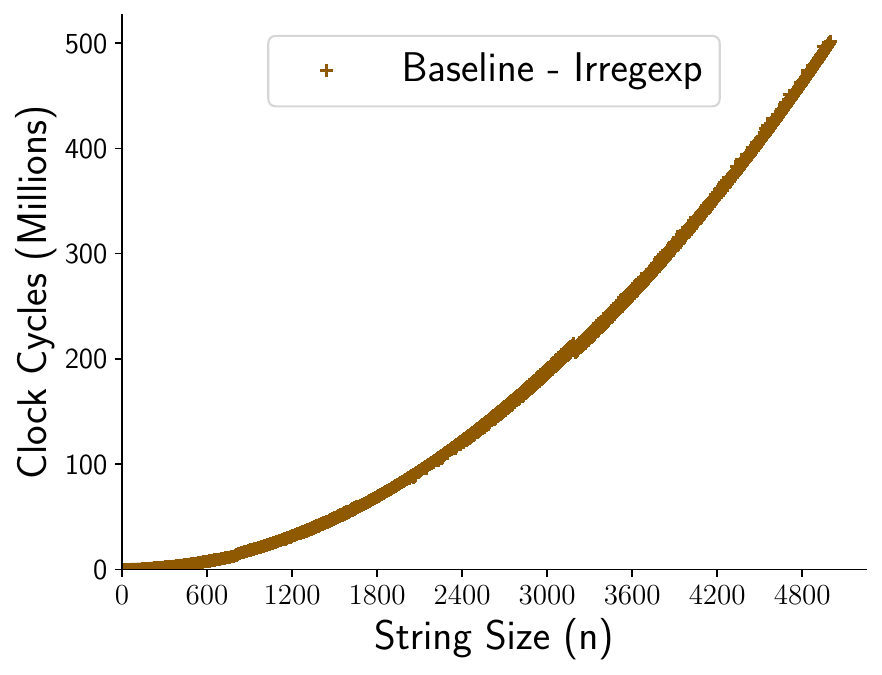}%
  \vspace{-0.3cm}%
  \caption{Lookarounds String Complexity}
  \label{fig:bench_lastr}
  \end{minipage}
\end{figure}


Previous sections provide theoretical arguments to support our algorithmic-complexity claims.
Here, we exemplify these claims on specific families of regexes and strings to confirm that, in practice, our implementations verify the following:
\textbf{(C1)} our capture reset algorithm is linear in $\size{r}$ when the previous V8Linear algorithm is quadratic;
\textbf{(C2)} our nonnullable plus construction is linear in $\size{r}$ when the previous V8Linear algorithm is exponential;
\textbf{(C3)} our CIN/CDN plus algorithm is linear in $\size{r}$ when the previous V8Linear algorithm is exponential;
\textbf{(C4)} our captureless lookbehind algorithm is linear in both $\size{r}$ and $\size{s}$ while backtracking is not linear in $\size{s}$;
\textbf{(C5)} our unrestricted lookaround algorithm is linear in both $\size{r}$ and $\size{s}$ while backtracking is not linear in $\size{s}$.
We compare the following configurations when applicable:
[OCaml] our prototype OCaml engine with the linked list implementation of Section~\ref{sec:tradeoff};
[oldV8L] V8Linear before our changes;
[newV8L] V8Linear after our changes; and
[Irregexp] the V8 backtracking engine (with compilation to native code enabled).
We modified V8Linear to allow any number of nested plus.
For all engines, measurements are done with the \texttt{rdtsc} instruction to estimate CPU cycles.
We measure 10 repetitions of \texttt{match()} after 10 warmup repetitions and take the median of 5 measurements.
We do not measure bytecode compilation.

\textbf{(C1) Capture Reset}
We evaluate the regex family $r_0=\regex{a}$, $r_{n+1}=\regex{($r_n$)*}$ on string \str{a$^{100}$}, a case where the bytecode grows quadratically if one uses the \bytecode{ClearReg} instructions of V8Linear.
We compare [OCaml] to [oldV8L] on Figure~\ref{fig:bench_capture} and confirm that our algorithm behaves linearly.
Note that here [oldV8L] suffers from two sources of regex-size quadraticity: the \bytecode{ClearReg} instructions and the array implementation (Section~\ref{sec:tradeoff}).

\textbf{(C2) NonNullable Plus}
We consider the regex family $r_0=\regex{a}$, $r_{n+1}=\regex{$r_n$+}$ on string \str{a$^{100}$}.
We compare [newV8L] to [oldV8L] on Figure~\ref{fig:bench_nn} and confirm that the construction of Figure~\ref{fig:usual_plus} avoids the exponential complexity of [oldV8L].

\textbf{(C3) Nullable Plus}
We consider the regex family $r_0=\regex{a|(\,\^\,)}$, $r_{n+1}=\regex{$r_n$+}$ on string \str{b}.
This is a worst case for our algorithm, since each plus is a CDN for which we need to compute the nullability.
After a match is found, each plus also needs to be executed again to reconstruct the capture group inside all of them.
We compare [OCaml] to [oldV8L] on Figure~\ref{fig:bench_cdn}.
Even in this worst case, our algorithm exhibits regex-size linearity.

\textbf{(C4) Captureless Lookbehinds}
To show linearity in $\size{s}$, we consider the regex \regex{b\noncap{a(?<=ba*)}*} and the string family $s_n=$\str{ba$^{n}$}.
On Figure~\ref{fig:bench_lookbehind}, we compare [newV8L] to the backtracking engine [Irregexp], since [oldV8L] did not support lookbehinds.
The backtracking engine exhibits quadratic complexity, since at each string position it tries to match the lookbehind and reads the string backward until the beginning.
Our algorithm does a single pass over the entire string.

\textbf{(C5) Lookarounds}
To show linearity in $\size{r}$, we first consider the regex family
$r_0=\regex{(a*)b}$, $r_{n+1}=\regex{a(?=$r_n$)}$ on string \str{a$^{1000}$b} on Figure~\ref{fig:bench_lareg}.
This is a worst case for our algorithm, since each nested lookahead needs to be run in the third step to reconstruct the capture group inside.
However, our algorithm still exhibits regex-size linearity, like the backtracking algorithm which does not even need to backrtack in this particular case.
To show linearity in $\size{s}$ in Figure~\ref{fig:bench_lastr}, we consider the regex \regex{c\noncap{a(?=a*(?<=c(a*))b)}*} on the string family $s_n=$\str{ca$^{n}$b} and compare [OCaml] to [Irregexp].
Since there are two lookarounds, three passes over the entire string are enough to find a match and our engine exhibits string-size linearity.

\section{Related Work}
\label{sec:related}

\parbf{Lookarounds}
While captureless lookarounds are reminiscent of regex intersection, it has been shown that constructing the NFA for a regex with one or more intersections comes with an exponential size increase in some cases~\cite{succinctness}.
Both lookbehinds and lookaheads are commonly known to only be supported by backtracking engines~\cite{re2_unsupported}.
Recent work has managed to integrate lookarounds into a derivative-based engine in .NET7~\cite{dotnet_pldi}.
When matching a lookaround, their algorithm interrupts the matching of the main expression to start a new engine on the lookaround.
Consequently, nesting lookarounds in quantifiers results in polynomial complexity, with the exponent increasing with nesting.
Even without lookarounds, their engine achieves linearity in $\size{s}$ but not in $\size{r}$.
We achieve linearity in both, even in the presence of lookarounds.
Another work~\cite{selective_memo} has shown that memoized backtracking engines can support captureless lookaheads in linear time, but with an additional memory overhead $\bigo{\size{r}\times\size{s}}$.
In contrast, our captureless lookbehind algorithm has no memory overhead.
Our unrestricted lookaround algorithm also has a similar space complexity, but it allows captures inside lookarounds for the first time.

Our captureless lookbehind algorithm was inspired by two observations from related work.
First, lookaheads can be encoded with Alternating Finite Automata (AFAs)~\cite{rewla}.
Second, AFAs can represent LTL formulas and be model-checked linearly if one reads the string (or the LTL \textit{trace}) backwards and reverses the direction of the automaton~\cite{afa_mc}.
Reversing the regex and reading the string backwards is usually incompatible with capture priority, but captureless lookbehinds lend themselves well to this treatment.
These two observations led to the development of our lookaround matching algorithms and to their first public description, in the issue tracker of V8\footnote{\url{https://bugs.chromium.org/p/v8/issues/detail?id=14099}}.
Later, related work independently proposed a similar algorithm to match regexes with lookarounds~\cite{efficient_lookarounds}.
In essence, that algorithm is a simplified version of the first two phases of our algorithm in Section~\ref{sec:lookarounds}. Just like in our first phase, an oracle is built by reversing regexes to indicate the positions at which each lookaround holds, but unlike our algorithms there is no support for capture groups (neither inside lookarounds nor in the main regex).
Their optimization to reduce memory usage for lookbehinds is a special case of the one we present in Section~\ref{sec:memoryless} when the main expression does not contain capture groups. The corresponding optimization for lookaheads is not directly applicable when the main regex contains capture groups (reversing the regex does not preserve capture group priority, so we do not apply any such transformation).

\parbf{Mitigating ReDoS}
ReDoS is a serious security concern for many applications~\cite{freezing_the_web,redos_impact}.
There exist two main kinds of mitigations.
The first one consists in \textit{detecting} regexes for which exponential backtracking can happen~\cite{revealer,static_dos,nfa_ambiguity,rescue,rexploiter,rat}, and \textit{repairing} them when possible (generating equivalent regexes without backtracking explosion)~\cite{repair_dos,repair_extraction,evil_harmless}.
The other consists in only using linear engines and avoid catastrophic backtracking altogether; this has become the default practice in Rust, Go and for anyone using the RE2 library.
By adding lookarounds to linear engines, our work extends the applicability of this second ReDoS mitigation.
Even simple regexes where backtracking is exponential because of other constructs (like \regex{\noncap{a*}*(?=b)} on a string of \str{a}s) can now be supported by linear engines.


\parbf{Other Regex Features}
Recent work~\cite{countingsetautomata,synchro_counting,bitvector} has explored matching counted repetition with a time complexity independent of the counters.
However, these solutions do not preserve the priority between threads that is needed to support capture groups.
Other work~\cite{polynomial_backref} has defined subsets of regexes with backreferences that can be matched in polynomial time.
There exist multiple semantics for backreferences, and~\cite{backref_reexamined} explored their relative expressive powers.  
\cite{posix_extraction} has presented algorithms to handle capture groups in longest-match POSIX semantics.

\section{Conclusion}
\label{sec:conclusion}

We have presented novel algorithms for matching most JavaScript regex linearly in the sizes of both the regex and the input string.
These guarantees are crucial for many applications working with user-provided regexes or strings: without them, applications are susceptible to regex-based denial-of-service attacks.
In the process, we have highlighted several incorrect assumptions in state-of-the-art linear engines, affecting both complexity and correctness.
We have presented the first algorithm to match unrestricted lookarounds in a linear-time engine, and thereby reduced the expressivity gap between backtracking and linear approaches.  Parts of our work, including our captureless lookbehind algorithm, are applicable to other linear regex engines, and could be included in Rust or RE2.
We have implemented and experimentally validated the practicality of all our algorithms, and merged some of them in the V8 JavaScript engine, putting them in the hands of millions of developers and making it more practical to chose secure-by-default execution.

Our work sheds light on the importance of seemingly minor semantic design choices.
For instance, JavaScript's semantics for nullable quantifiers requires more complex algorithms than other regex languages.
On the contrary, its capture-reset property enables us to support, for the first time in any regex language, linear-time matching of capturing lookarounds.

Because of backreferences, JavaScript regexes cannot be fully supported by linear engines.
The current trend in modern languages is to move away from backreferences and provide secure regex matching, either by default (Rust and Go) or as an alternative (.NET~\cite{dotnet_pldi}).
Our work shows how to bring these benefits to JavaScript, a language particularly affected by ReDoS~\cite{freezing_the_web}.
We show that this requires sacrificing very few features (backreferences, counted repetition, lazy nullable plus).
As a future direction, we note that the surprising amount of non-linearities and semantic subtleties that we uncovered suggests that JavaScript regexes would strongly benefit from a systematic formalization of the regex language and its engines.

\section*{Acknowledgments}
We thank Ludovic Mermod for working with us on implementing our algorithms in the V8 engine.
We thank V8 developers Patrick Thier, Jakob Linke and Olivier Flückiger for helpful discussions and code reviews.
We also thank Nate Foster and Viktor Kuncak for the feedback on this paper.

\section*{Artifact Availability}
A virtual machine image with our implementations is available as an artifact~\cite{artifact}.
It contains our OCaml implementations, our V8 patches, scripts to reproduce our benchmarks and analyses, and a script to rebuild and provision the virtual machine image itself.
The OCaml engine implementing our algorithms is also available online: \url{https://github.com/epfl-systemf/RegElk}.

\bibliographystyle{ACM-Reference-Format}
\bibliography{main}

\newpage

\appendix

\section{Example: Executing our Lookarounds Algorithm}
\label{sec:lookarounds_example}


Consider the regex $r=$\regex{(c)\noncap{a(?=a*(?<=c(a*))b)}*} on string \str{caab}.
This regex gets annotated as follows: \regex{(c)\grp{1}\noncap{a(?=\lkr{1}a*(?<=\lkr{2}c(a*)\grp{2})b)}*}.

\subsection{First step - Building the oracle}
\begin{wrapfigure}{l}{5cm}
  \scriptsize
\begin{tabular}{c}
\begin{lstlisting}[style=byt]
0: Fork 3 1
1: ConsumeAny
2: Jump 0
3: Consume c
4: Fork 5 7
5: Consume a
6: Jump 4
7: WriteOracle 2
\end{lstlisting}
\end{tabular}
\caption{Oracle Bytecode for $r\lkr{2}$}
\label{fig:ex_oracle2}
\end{wrapfigure}

In our first step, we build the oracle.
The oracle has a two rows (there are two lookarounds) and 5 columns (the string has 4 characters).
For the row of lookaround~\looka{2}, we consider $r\lkr{2}=$\regex{c(a*)}.
In that step, we remove the capture group and \bytecode{SetQuant} instructions, use a \bytecode{WriteOracle} instruction and add a \rgx{.*?} prefix.
The resulting bytecode is shown on Figure~\ref{fig:ex_oracle2}.
The first three instructions are for the lazy prefix.
Executing an NFA simulation, in the forward direction, on string \str{caab} finds matches at positions 1, 2 and 3 (every position between the \str{c} and the \str{b}).
\newline

\begin{wrapfigure}{L}{5cm}
  \scriptsize
\begin{tabular}{c}
\begin{lstlisting}[style=byt]
0: Fork 3 1
1: ConsumeAny
2: Jump 0
3: Consume b
4: CheckOracle 2
5: Fork 6 8
6: Consume a
7: Jump 5
8: WriteOracle 1
\end{lstlisting}
\end{tabular}
\caption{Oracle Bytecode for $r\lkr{1}$}
\label{fig:ex_oracle1}
\end{wrapfigure}

For the first row of the oracle (lookaround~\looka{1}), we consider $r\lkr{1}=$\regex{a*(?<=\lkr{2})b}.
Since this is a lookahead, we compute $\mathit{rev}(r\lkr{1})=$\regex{b(?<=\lkr{2})a*} and add a \rgx{.*?} prefix.
The resulting bytecode is shown on Figure~\ref{fig:ex_oracle1}.
The inner lookaround~\looka{2} is simply compiled to a single \bytecode{CheckOracle} instruction which reads the previous row of the oracle we just computed before.
Executing an NFA simulation, in the backward direction, on string \str{caab} also finds matches at positions 1, 2 and 3.

\def\oraclef{\text{\sffamily X}}
\def\oraclet{\checkmark}

The final oracle is shown on the following table:
\begin{tabular}{c|c c c c c}
  String Position & 0 & 1 & 2 & 3 & 4\\
  \hline
  $r\lkr{1}$ & \oraclef & \oraclet & \oraclet & \oraclet & \oraclef\\
  $r\lkr{2}$ & \oraclef & \oraclet & \oraclet & \oraclet & \oraclef\\
\end{tabular}
\newline

\subsection{Second step - Matching the main expression and groups \group{0} and \group{1}}
\begin{wrapfigure}{L}{6.6cm}
  \scriptsize
  \begin{tabular}{c}
\begin{lstlisting}[style=byt]
0: Fork 3 1
1: ConsumeAny
2: Jump 0
3: SetReg #0:entry
4: SetReg #1:entry
5: Consume c
6: SetReg #1:exit
7: Fork 8 12
8: SetQuant 1
9: Consume a
10: CheckOracle 1
11: Jump 7
12: SetReg #0:exit
13: Accept
\end{lstlisting}
  \end{tabular}
  \caption{Bytecode of $r$}
  \label{fig:ex_main}
\end{wrapfigure}

In the second step, we compile and match the main expression.
Note that the entire lookahead will be compiled to a single bytecode instruction \bytecode{CheckOracle 1}.
We compile the regex \newline\regex{.*?((c)\grp{1}\noncap{a(?=\lkr{1}a*(?<=\lkr{2}c(a*)\grp{2})b)}*)\grp{0}}\newline
and show its bytecode on Figure~\ref{fig:ex_main}.

Executing an NFA simulation forward on this expression will find a best match where group\group{0} is \str{caa}, group\group{1} is \str{c} and group\group{2} is undefined.
The values for groups \group{0} and \group{1} are correct because these groups are defined inside the main expression.
However, group\group{2} is defined inside a lookahead.
\newline
\newline

\subsection{Final step - Getting the value of group\group{2}}

\begin{wrapfigure}{L}{6cm}
  \scriptsize
  \begin{tabular}{c}
\begin{lstlisting}[style=byt]
0: Fork 1 4
1: SetQuant 2
2: Consume a
3: Jump 0
4: CheckOracle 2
5: Consume b
6: Accept
\end{lstlisting}
  \end{tabular}
  \caption{Reconstruction Bytecode of $r\lkr{1}$}
  \label{fig:ex_reconstruct1}
\end{wrapfigure}

We now reconstruct the value of group\group{2}.
We can see that in the match of the second step, the winning thread last used the oracle for lookahead~\looka{1} at position 3.
We compile the lookahead $r\lkr{1}$ to bytecode.
This time, we do not reverse it and keep capture groups.
Its bytecode is shown on Figure~\ref{fig:ex_reconstruct1}.
We start the NFA simulation in a forward direction at position 3.
A match is found, where the lookbehind~\looka{2} was last used at position 3, meaning that we need to run it too in order to get the value of the group\group{2}.
\newline

\begin{wrapfigure}{l}{6cm}
  \scriptsize
  \begin{tabular}{c}
\begin{lstlisting}[style=byt]
0: Fork 1 6
1: SetQuant 3
2: SetReg #2:entry
3: Consume a
4: SetReg #2:exit
5: Jump 0
6: Consume c
7: Accept
\end{lstlisting}
  \end{tabular}
  \caption{Reconstruction Bytecode of $r\lkr{2}$}
  \label{fig:ex_reconstruct2}
\end{wrapfigure}

We reverse and compile $r\lkr{2}$ to bytecode, without removing capture groups.
Its bytecode is shown on Figure~\ref{fig:ex_reconstruct2}.
We start the NFA simuation in a backward direction, starting at position 3.
A match is found where capture group\group{2} has value \str{a}.
We now have the full results of matching $r$ on string \str{caab}:
group\group{0} has value \str{caa},
group\group{1} has value \str{c} and
group\group{2} has value \str{a}.

\end{document}